\documentclass[%
 reprint,
 amsmath,amssymb,
 aps,
]{revtex4-2}

\usepackage{graphicx}% Include figure files
\usepackage{dcolumn}% Align table columns on decimal point
\usepackage{bm}% bold math
\usepackage{empheq}
\usepackage{siunitx}
\DeclareSIUnit{\dBm}{dBm}
\usepackage{xcolor}

\usepackage{fixltx2e}
\usepackage{soul}
\usepackage{float}
\usepackage[normalem]{ulem}
\begin{document}

\preprint{APS/123-QED}

\title{Engineering \qty{}{\cm}-scale true push-pull electro-optic modulators in a suspended GaAs photonic integrated circuit platform by exploiting the orientation induced asymmetry of the Pockels $r_{41}$ coefficient}

\author{Haoyang Li}
\email{haoyang.li@bristol.ac.uk}
\affiliation{Quantum Engineering Technology Labs and Department of Electrical and Electronic Engineering, University of Bristol,
Woodland Road, Bristol BS8 1UB, United Kingdom}

\author{Robert Thomas}
\affiliation{Quantum Engineering Technology Labs and Department of Electrical and Electronic Engineering, University of Bristol,
Woodland Road, Bristol BS8 1UB, United Kingdom}

\author{Pisu Jiang}
\affiliation{Quantum Engineering Technology Labs and Department of Electrical and Electronic Engineering, University of Bristol,
Woodland Road, Bristol BS8 1UB, United Kingdom}

\author{Krishna C. Balram}
\email{krishna.coimbatorebalram@bristol.ac.uk}
\affiliation{Quantum Engineering Technology Labs and Department of Electrical and Electronic Engineering, University of Bristol,
Woodland Road, Bristol BS8 1UB, United Kingdom}

\date{\today}% It is always \today, today,

\begin{abstract}
Electro-optic modulators (EOMs) underpin a wide range of critical applications in both classical and quantum information processing. While traditionally the focus has been on building these devices in materials with large Pockels coefficient (mainly ferroelectric insulators like lithium niobate), there is a need to engineer EOMs in a semiconductor platform with a view towards device stability (in radiation-hard environments), manufacturability (wafer size and foundry compatibility) and integration (with active electronics and quantum confined structures). Here, we demonstrate true push-pull EOMs in a suspended GaAs photonic integrated circuit (PIC) platform by exploiting the orientation induced asymmetry of the Pockels $r_{41}$ coefficient, and folding the two arms of a \qty{}{\cm}-scale Mach-Zehnder interferometer (MZI) modulator along two orthogonal crystal axes. Our work also shows the potential of incorporating ideas from micro-electro-mechanical systems (MEMS) in integrated photonics by demonstrating high-performance active devices built around \qty{}{cm}-scale suspended waveguides with sub-\qty{}{\um} optical mode confinement.  
\end{abstract}

\maketitle

%\tableofcontents

\section{Introduction}

Electro-optic modulators (EOMs) are critical for mapping analog and digital signals from the microwave to the optical domain for a wide range of applications in both classical and quantum information processing. These span from developing transceivers for fiber-optic communication systems \cite{wooten2000review, zhang2021integrated} to radio-over-fiber applications in microwave photonics \cite{capmany2007microwave}. Recently, their performance (propagation loss and electro-optic coupling strength) has been improved to the point that they are leading candidates for building efficient microwave to optical photon transducers \cite{han2021microwave, balram2022piezoelectric}, despite the $\approx 10^5\times$ difference between the wavelengths of the fields involved (\qty{}{cm} for the microwave, \qty{}{\um} for the optical) \cite{holzgrafe2020cavity, mckenna2020cryogenic, warner2023coherent}.

Both historically and recently \cite{wooten2000review, zhang2021integrated}, state-of-the-art EOMs have been built around ferroelectric insulators \cite{maldonado1995electro} like lithium niobate \cite{zhang2021integrated}, lithium tantalate \cite{wang2024ultrabroadband} and barium titanate \cite{abel2019large, alexander2024manufacturable} due to their high Pockels coefficient and low intrinsic optical absorption. On the other hand, ferroelectric insulators have certain intrinsic material limitations. These include long-term stability exemplified by the relaxation of the electro-optic response \cite{holzgrafe2020cavity} and the resulting DC bias drift \cite{wooten2000review, wang2024ultrabroadband}, and inertness to reactive ion etching chemistries. The reliance on Ar-ion based physical etching techniques, with extensive sidewall redeposition and waveguide sidewall angles $\approx \qty{60}{\degree}$ \cite{kaufmann2023redeposition} makes it difficult to leverage photonic bandgap structures \cite{meade2008photonic} to shape and control waveguide dispersion \cite{kawahara2024high, nardi2024integrated}. If we further desire that the material platform build on and leverage existing infrastructure investments in microelectronics \cite{wang2024ultrabroadband} with a view towards scalability, integration with active electronics and long-term unit economic costs, then the choice can not be made based purely on device metrics. This is best illustrated by the fact that modern data centres rely heavily on silicon photonics based transceivers \cite{glick2023integrated}, even though their individual device performance lags far behind state-of-the-art lithium niobate (LN) devices.

These factors make it interesting to continuously push the performance of EOMs fabricated in semiconductor platforms, in complement to efforts on ferroelectric insulators. Indium phosphide (InP) has been the traditional material of choice mainly due to the prospect of being able to monolithically integrate lasers on the same die \cite{hoefler2019foundry}. But, if we take the silicon photonics example above and consider the question of leveraging existing infrastructure investments, we argue that gallium arsenide (GaAs) presents a more logical choice to make the silicon-like electronics to photonics manufacturing leap by building on existing GaAs foundry investments \cite{wang2014gaas}. GaAs EOMs have a long and distinguished history \cite{dagli1999wide, walker2013optimized} and have found a niche in space-based (satellite) applications \cite{walker201950ghz} where GaAs' radiation hardness and space qualification (from the electronics side) gives it a significant advantage over other material platforms.

In addition to potential (electronic) foundry compatibility, another major driver for the pursuit of efficient GaAs EOMs is that the Ga(Al,In)As material system is the most extensively studied and well-developed for hosting quantum confined structures, in particular quantum dots and wells. InAs based quantum dots \cite{lodahl2015interfacing} hosted in a GaAs lattice currently provide the brightest solid-state single photon sources \cite{tomm2021bright}, and are currently the leading candidate for generating cluster states \cite{scott2022timing, alexander2024manufacturable} necessary for photonic implementations of measurement based quantum computing (MBQC). Implementing feedforward operations \cite{scott2022timing, alexander2024manufacturable} in MBQC architectures places a premium on integrated high-performance EOMs. 

\begin{figure*}[t]
    \centering
    \includegraphics[width=\textwidth]{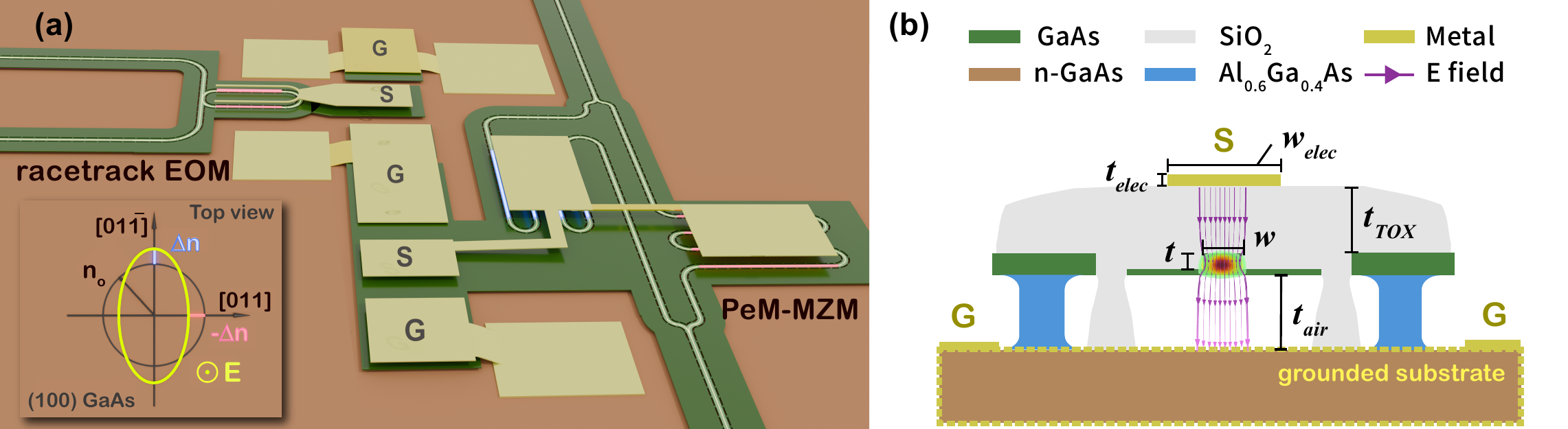}
    \caption{(a) Schematic view of the suspended GaAs PIC platform showing the perpendicularly meandering Mach-Zehnder modulator (PeM-MZM, bottom) and racetrack resonator based EOM (top) on a (100) oriented GaAs wafer, showing the relative position between the electrodes and the underlying waveguides. The inset shows the planar projection of the GaAs index ellipsoid. Without an applied electric field along the $[100]$ axis, GaAs is optically isotropic in-plane (black circle). When an external electric field is applied along the $[100]$ axis, the ellipsoid deforms (yellow ellipse) with major and minor axes along the $[011]$ or $[01\bar{1}]$ directions. Key for the PeM-MZM push-pull operation is that the refractive index change is equal and opposite in the two directions. (b) 2D cross section of suspended GaAs rib waveguide showing the interaction between the propagating optical field (transverse electric mode field calculated using FEM is overlaid to scale) and the out of plane DC / RF field (purple streamlines). Device parameters used in the simulations: waveguide width $\textit{w} = \qty{540}{\nm}$, rib etch depth $t = \qty{240}{\nm}$, top oxide thickness $t_\text{TOX} = \qty{2.2}{\um}$, Al\textsubscript{0.6}Ga\textsubscript{0.4}As/air gap thickness $t_\text{air} = \qty{2}{\um}$,  electrode thickness $t_\text{elec} = \qty{460}{nm}$, top electrode width $w_\text{elec} = \qty{5}{\um}$. The different components in the device are shown in the legend. The linear EO effect induces a refractive index change of $\Delta n_\text{eff} = $\qty{1.279e-6}{\per\volt} in the GaAs waveguide due to the applied electric field.}
    \label{fig:schematic}
\end{figure*}

Despite their long development history, GaAs based EOMs have shared some common themes. They have generally relied on vertical epitaxially grown p-i-n diodes \cite{walker2013optimized} which are reverse biased for the EO effect. To reduce free carrier absorption and also to account for the weak index contrast between GaAs core and AlGaAs cladding layers (${\Delta}n \approx 0.2$), the mode sizes are typically $\approx$ \qty{3}{\um} and the bend radii $>$ \qty{100}{\um} which limits the component density. Given that the refractive index of GaAs is comparable to Si at telecommunication wavelengths \cite{jiang2020suspended}, one should ideally be able to get silicon-like component density with the added benefit of high-performance EOMs by increasing the index contrast, either via suspension \cite{khurana2022piezo} or by working with a gallium arsenide on insulator platform using either wafer bonding \cite{stanton2020efficient} or membrane transfer \cite{roelkens2024present}. The question of whether to use suspensions or wafer bonding to build high-performance GaAs devices is an open one and in many ways mirrors the debate in the LN EOM community \cite{mookherjea2023thin}. We take the view that if bonding (and substrate removal) can be avoided without compromising device performance \cite{thomas2023quantifying} and reliability, then one should do so. Moreover, suspended platforms (and incorporating MEMS-based approaches) have natural advantages whenever opto-mechanical interactions are involved, such as in building microwave to optical quantum transducers \cite{balram2022piezoelectric} using acoustics \cite{khurana2022piezo} as an intermediary.  

We illustrate the benefits of strong (sub-\qty{}{\um}) confinement and the resultant reduction in device footprint by demonstrating true \qty{}{\cm}-scale push-pull modulators in GaAs. To clarify, by \textit{true}, here we are referring to modulators analogous to X-cut LN \cite{zhang2021integrated}, wherein the same voltage is applied to the two arms of the phase modulator, configured as a Mach-Zehnder interferometer, but one gets equal and opposite phase shifts. Unlike the X-cut LN case, which relies on lateral (in-plane) fields by exploiting the Pockels $r_{33}$ coefficient and allows the signal electrode to be located at the centre of two outer ground planes, in GaAs, the Pockels $r_{41}$ coefficient requires a vertically oriented field (as illustrated in Fig.\ref{fig:schematic}(a,b)) which results in equal phase shifts in the two parallel MZI arms. To build an EOM, therefore, one needs to apply RF signals anti-phase to the two MZI arms in a centre-tapped configuration (series push-pull) which requires additional bias and DC-decoupling circuitry \cite{walker1987high}. To work around the issue in GaAs \cite{dagli_high-speed_2007_ch4, midolo2017electro}, we use the fact that the application of a vertical electric field (along the $[100]$, z-axis) breaks the in-plane refractive index symmetry. Light that is propagating along the $[011]$ crystal axis picks up an equal and opposite phase shift to that propagating along the $[01\bar{1}]$ axis (assuming transverse electric polarization, TE mode).

This is illustrated by the (in-plane) index ellipsoid shown in the inset of Fig.\ref{fig:schematic}(a) for one polarity of the vertical electric field. The ellipsoid will flip from being oblate to prolate as the field switches polarity. By folding the waveguide in the two arms of the MZI to lie (predominantly, ignoring the bends) along the $[011]$ and $[01\bar{1}]$ axes respectively, one achieves equal and opposite phase shifts in the two arms. This design is enabled primarily by the strong index contrast (${\Delta}n\approx$ 2) enabled by waveguide suspension, which allows tight folding, while maintaining a compact on-chip footprint. Building high-performance EOMs while working with the low $r_{41}$ coefficient of GaAs requires \qty{}{cm}-scale arm lengths, which we demonstrate below, showing how far MEMS based ideas can be used to push integrated photonics platforms.

\section{Device design and fabrication}

\begin{figure*}[t] 
    \centering
            \includegraphics[width=\textwidth]{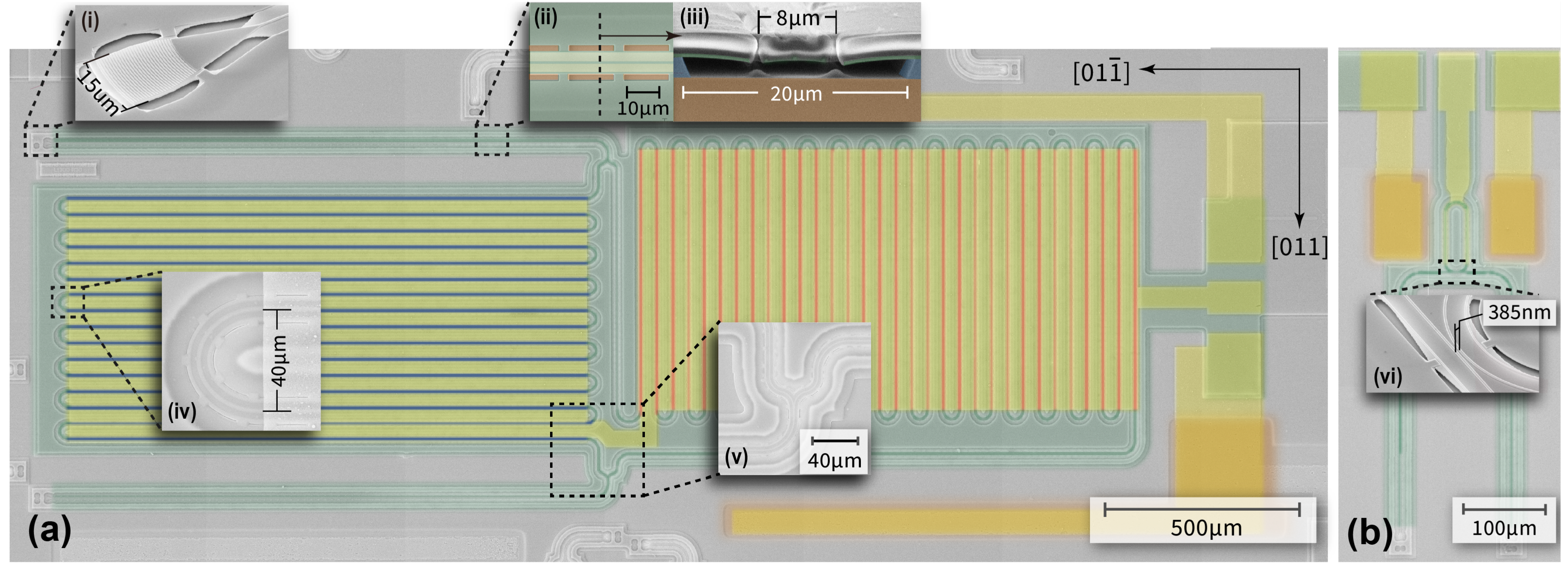}
  \caption{False-colored SEM view of suspended GaAs (a) PeM-MZM and (b) racetrack EOM devices. The electrodes (yellow) covers the GaAs waveguide (green) with MZ arms meandering along $[011]$ (red highlight) and $[01\bar{1}]$ (blue highlight) directions. The uncolored regions represents GaAs substrate covered by deposited SiO\textsubscript{2} layer. Etch windows on SiO\textsubscript{2} layer are opened adjacent to the devices exposing bottom doped substrate (orange), allowing ground electrodes to form ohmic contact with substrate. Insets (i-vi) show zoomed views of key individual device components making up the EOM: (i) \qty{15}{\um}-wide surface-normal grating coupler, (ii) top view of the rib waveguide suspended by \qty{19}{\um} spaced tethers, (iii) rib waveguide cross section showing a \qty{20}{\um}-wide air gap opened beneath the waveguide, (iv) Euler U-shape bend with bend width of \qty{40}{\um} to mitigate bending loss, (v) 1-to-2 Y-splitter, (vi) bus waveguide-resonator coupler for the racetrack EOM with a coupling gap of \qty{385}{nm}. SEM insets (i, ii, vi) are taken before capping the oxide, to give a clearer view of optical components and their suspension.}
  \label{fig:SEM}
\end{figure*}

Fig.\ref{fig:schematic}(a) shows a schematic of our proposed devices. The perpendicularly meandering Mach-Zehnder modulator (PeM-MZM) with the two waveguide arms oriented along the $[011]$ and $[01\bar{1}]$ respectively is indicated. Application of a vertical electric field (an FEM simulation of the electric field lines are shown in Fig.\ref{fig:schematic}(b)) breaks the in-plane refractive index symmetry and the (in-plane) index ellipsoid is oriented as shown in the figure inset. Given that GaAs is a zinc-blende crystal with symmetry group ($\bar{4}3m$), the change in refractive index ($\Delta$n) due to the linear electro-optic effect using the Pockels $r_{41}$ coefficient, under the action of a vertically applied electric field can be written as:   

\begin{equation}
    {\Delta}n_{[011]} = +\frac{1}{2}n_o^3r_{41}E_{\perp,[100]} 
\end{equation}

\begin{equation}
    {\Delta}n_{[01\bar{1}]} =  -\frac{1}{2}n_o^3r_{41}E_{\perp,[100]} 
\end{equation}

where $n_o$ is the GaAs refractive index (3.37 at \qty{1550}{\nm}), $r_{41}$ = -\qty{1.5}{\pm \per\volt} is the relevant Pockels coefficient for the electro-optic interaction with a transverse electric (TE) polarized optical mode in the waveguide and a vertically oriented ($E_{\perp,[100]}$) electric field (either DC or RF). The equal and opposite signs of the refractive index change along the two crystal axes lies at the heart of the push-pull effect exploited in the PeM-MZM device. There is an additional quadratic EO effect, which is both significantly smaller, but more importantly gives equal phase shifts in the two arms, hence cancels out in this differential scheme. In theory, for the same applied electric field strength at the waveguide location, the refractive index change for GaAs based devices is $\approx 5 \times$ smaller than equivalent LN devices. To calibrate the push-pull effect and quantify the field strengths in the suspended waveguide platform, we also fabricate racetrack microring resonator based EOMs in the same platform where the sides of the racetrack are oriented along the crystal axes as shown in Fig.\ref{fig:schematic}(a), although here the quadratic EO contribution doesn't cancel out.

The devices are fabricated on a \qty{340}{\nm} GaAs membrane which is released by undercutting an underlying Al$_{0.6}$Ga$_{0.4}$As buffer layer using hydrofluoric acid (HF). The fabrication of the GaAs PIC follows a process similar to our previous work \cite{jiang2020suspended, khurana2022piezo, thomas2023quantifying}. The suspended waveguide platform is encapsulated in silicon oxide deposited by plasma enhanced chemical vapor deposition. The oxide locks the structure mechanically providing rigidity \cite{jiang2020suspended}, and also serves to offset the signal electrode from the waveguide layer (cf. Fig.\ref{fig:schematic}(b)). To build EOMs, we open up windows in the oxide layer to define the signal and ground electrodes and define the contacts using lift-off with an additional aligned lithography step. The $r_{41}$ coefficient requires a vertically oriented electric field for operation. Therefore, the signal contact is deposited on top of the waveguide (offset by the oxide thickness $\approx$ \qty{2}{\um}). To get the bottom contact underneath the waveguide to maximize the verticality of the dropped RF field (see Fig.\ref{fig:schematic}(b) for an FEM simulation showing the electric field lines around the waveguide), we use an n-doped GaAs substrate (\qty{1e18}{\per\cubic\cm}) and use an annealed AuGe/Ni/Au metal stack to get an ohmic contact, see Appendix \ref{sec:appendix-fab} for further details.

Fig.\ref{fig:SEM}(a,b) show false-colored SEM images of the PeM-MZM and racetrack EOM devices respectively. The different components of the device are shown by zoomed-in images added to the figure inset. Light is coupled onto and off the chip using focusing grating couplers (Fig.\ref{fig:SEM}(i)) and routed using suspended rib waveguides (Fig.\ref{fig:SEM}(ii, iii)). For the PeM-MZM designs, we split the light into the two MZ arms at the input using a Y-coupler (Fig.\ref{fig:SEM}(v)) and we use an identical Y-coupler at the output to recombine the light from the two arms. The push-pull effect originates from the orientation of the waveguide arms along two orthogonal axes as shown in the figure. The high refractive index contrast and strong mode confinement allows us to tightly fold the MZM. We use Euler bends \cite{cherchi2013dramatic} with effective bend radii of \qty{20}{\um} (Fig.\ref{fig:SEM}(iv)) to ensure minimal mode mismatch between the straight and bent waveguide regions.

The PeM-MZM shown in Fig.\ref{fig:SEM}(a) are designed with arm lengths of \qty{2.5}{\cm} and \qty{2.36}{\cm} for the beam paths oriented along the $[011]$ and the $[01\bar{1}]$ axes respectively. We work with an asymmetric MZI design in these first-generation devices as it helps ease constraints on the layout and the spectral dependence on transmission helps us bound the losses of internal components like grating couplers, bends and Y-splitters. The overall design takes up an on-chip footprint of \qty{1}{\mm} $\times$ \qty{3.1}{\mm}. We were conservative in our designs with respect to lateral undercut provision and the radii of the Euler bends to ensure working devices in these first generation experiments. By optimizing both parameters, we expect to see a further $2-5\times$ reduction in device footprint. 

The scale of the device in Fig.\ref{fig:SEM}(a) clearly shows the potential of incorporating MEMS based techniques into integrated photonics platforms \cite{quack2019mems}, beyond silicon wherein thin film on-insulator substrates are not readily available or are limited in substrate size. We maintain sub-\qty{}{\um} mode confinement over \qty{2.5}{\cm} scale on-chip path lengths, and the platform is stable to enable sensitive on-chip interferometry. To ease the fabrication constraints in these proof-of-principle devices, we chose to work with lumped electrodes for the EOMs, shown schematically in Fig.\ref{fig:schematic}(a), and indicated by the gold pads in Fig.\ref{fig:SEM}(a,b) for the PeM-MZM and the racetrack EOM respectively. For the PeM-MZM device in Fig.\ref{fig:SEM}(a), the electrode overlaps \qty{2.08}{\cm} of the folded waveguide in both arms to maintain the symmetry of the push-pull operation. 

\section{Device characterization}

\begin{figure*}[th]
\centering
    \includegraphics[width=\linewidth]{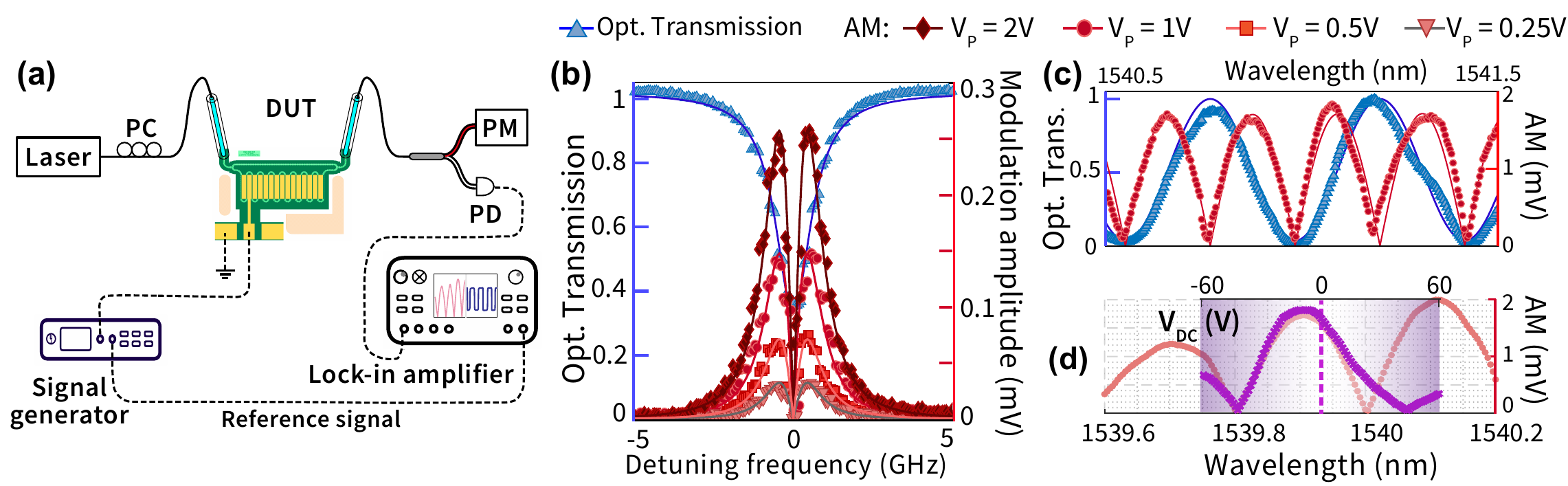}
    \caption{\label{fig:EOmod} (a) Experimental setup used for electro-optic modulation characterization. (b) A representative mode from the normalized optical transmission spectrum (blue) of the racetrack EOM, showing a loaded quality factor $Q \sim$ \qty{1.47e5}{} and extinction depth $ER = \qty{4.83}{\dB}$ (Appendix \ref{sec:appendix-opt}). The measured spectrum is fit using a Lorentzian lineshape (blue, solid). The measured modulation amplitude (lock-in signal) is shown (red, scatter) and the predicted fit is shown in shades of red for different applied modulation voltages ranging from \qty{0.25}{\volt} to \qty{2}{\volt}. We can see that the measured AM signal is clearly linear within this range. (c) A representative section of the optical transmission spectrum and modulation amplitude spectrum of the PeM-MZM device from Fig.\ref{fig:SEM}(a). The optical spectrum (blue scatter, normalized) is fitted with a sinusoidal curve (blue line), while the AM spectrum (red scatter) is fitted with a half-wave rectified sinusoidal model (red line). See Appendix \ref{sec:appendix-param} for details on the fitting procedure. (d) AM spectrum of the PeM-MZM device measured with \qty{0}{\volt} DC bias and \qty{1}{\volt} RF voltage amplitude (pink scatter). Overlaid purple crosses show the AM response as the DC bias is swept from \qty{-60}{\volt} to \qty{+60}{\volt} (top x-axis) with the laser wavelength parked at the dashed line, and the RF signal amplitude fixed at \qty{1}{\volt}. We believe the non-alignment of the data near $\lambda$=\qty{1540}{\nm} is due to temperature induced spectral shifts during data acquisition.}
\label{fig:ss_mod}
\end{figure*}

We characterize linear electro-optic modulation in our devices using the setup shown in Fig.\ref{fig:ss_mod}(a). Light from a tunable laser (Santec, TSL-550) is coupled into and out of the device under test (DUT) from a fiber array using grating couplers. As the laser wavelength is scanned, a modulation (AC) signal of frequency \qty{1}{\MHz}, and peak amplitude \qty{1}{\volt} for PeM-MZM (\qty{0.25}{\volt}-\qty{2}{\volt} for the racetrack EOM) is applied to the ground-signal-ground electrode configuration using a microwave probe. The transmitted optical signal is measured using both an optical power meter (Thorlabs, PM100USB) to record the transmission spectrum, and with a high-speed photodiode (Optilab, APR-10-MC), whose output is fed into a lock-in amplifier (Stanford Research Systems, SR865A) for modulation amplitude measurement. The signal generator (Tektronix, AFG2021) provides the reference signal for the lock-in, as indicated in Fig.\ref{fig:ss_mod}(a). The phase modulation induced by the EO effect is translated to amplitude modulation (AM) by the spectral dependence of the DUT transmission, and this translated AM is recorded as the modulation amplitude by the lock-in amplifier from the photodiode output.

Fig.\ref{fig:ss_mod}(b,c) shows the measured modulation amplitude spectra overlaid on the optical transmission spectra for the racetrack EOM and the PeM-MZM devices respectively. The measured modulation amplitude as a function of laser wavelength agrees well with the gradient of the optical transmission spectra, in line with the PM to AM translation argument discussed above. Fitting the modulation amplitude (see Appendix \ref{sec:appendix-param} for details) allow us to extract the modulation efficiency, expressed as a spectral tunability ($\eta$, [\qty{}{\pm\per\volt}]) or an equivalent half-wave voltage ($V_{\pi}$) need to shift the transmission from a maxima to a minima (or vice-versa). For racetrack EOMs with a loaded quality factor $Q\approx$ \qty{1.47e5}{} and extinction ratio $ER = $ \qty{4.83}{\dB}, we extract an $\eta=$\qty{0.351 \pm 0.008}{\pm\per\volt} and a $V_{\pi}=$\qty{31.9 \pm 0.8}{\volt}. For the PeM-MZM devices, the values are $\eta=$\qty{0.139 \pm 0.003}{\pm\per\volt} and a $V_{\pi}= $\qty{54.3 \pm 1.3}{\volt}. The $V_\pi$ for PeM-MZM can also be directly quantified through a DC sweep measurement, as shown in Fig.\ref{fig:ss_mod}(d). Here, we repeat the modulation experiment as in Fig.\ref{fig:ss_mod}(a), but add a DC bias voltage on top of the AC voltage (amplitude = \qty{1}{\volt}). By sweeping the DC bias voltage, one can in principle traverse the optical transmission spectrum, as shown in Fig.\ref{fig:ss_mod}(d), and read out the $V_{\pi}$ directly. The racetrack EOM measurement serves as a reference for the more complex PeM-MZM devices. From the modulation measurements, we can extract an equivalent refractive index change per unit applied voltage for both devices. This gives us $\Delta n_\text{eff} = 1.084\times10^{-6}$\qty{}{\per\volt} for racetrack EOM and $\Delta n_\text{eff} = 6.97\times10^{-7}$\qty{}{\per\volt} for PeM-MZM. The extracted $\Delta n_\text{eff}$ for the racetrack EOM agrees well with the predicted $\Delta n_\text{eff} = 1.279\times10^{-6}$\qty{}{\per\volt} using FEM simulation (cf. Appendix \ref{sec:appendix-simu}). 

\begin{figure}[!htbp]
        \centering
        \includegraphics[width = \linewidth]{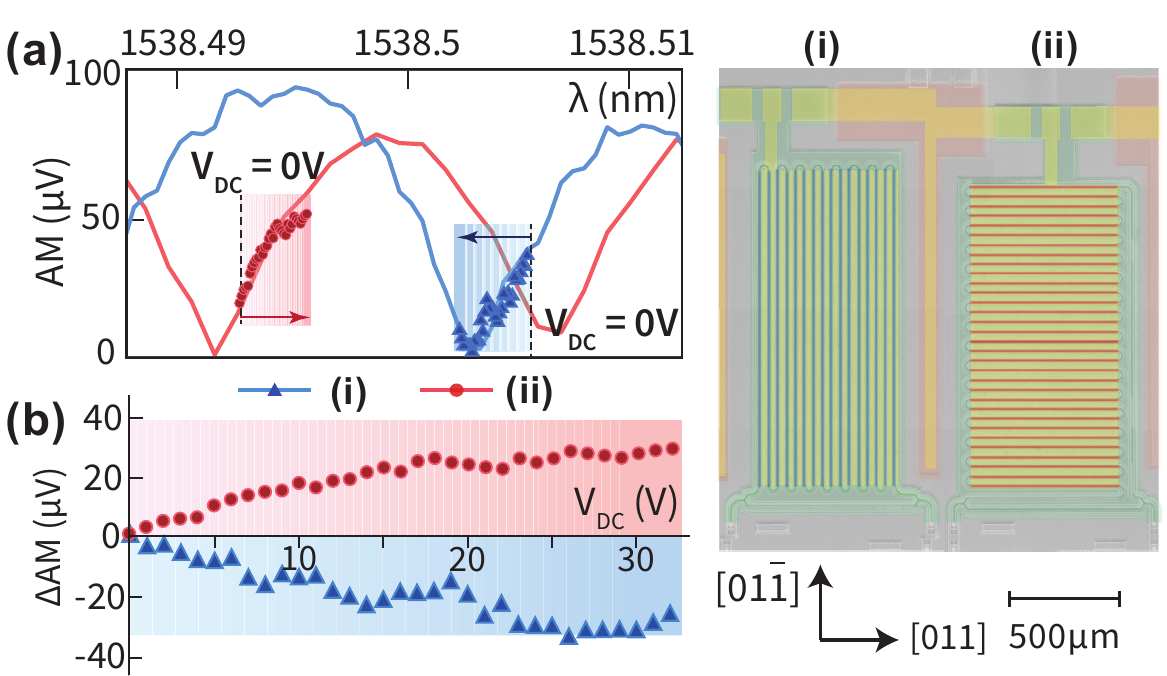}
    \caption{Control experiments to demonstrate the push-pull nature of the effect: DC bias induced phase shift on two SeM-MZM devices, with a single arm meandering along $[011]$(i) or $[01\bar{1}]$(ii) direction. The meandering arm lengths are designed to be nominally equal in the two cases. (a) AM spectra for SeM-MZMs driven by a \qty{1}{\MHz} modulation signal of amplitude \qty{1}{\volt}, (blue for (i), red solid for (ii)). Overlaid scattered plot (red circles and blue triangles) shows the shift in the AM spectrum when the DC bias voltage is swept from \qty{0}{\volt} (black dashed line) to \qty{32}{\volt}. The laser wavelength is indicated by the dashed line (b) Replotting the data from (a) to show the differential AM change as a function of applied DC bias voltage. The differential shift (${\Delta}$AM = AM($V_\text{DC}$)-AM(0)) is plotted with reference to the zero DC bias point. The opposite slopes of the differential AM voltage with respect to the bias voltage $V_{DC}$ clearly shows the push-pull effect in action.}
    \label{fig:DC_shift}
\end{figure}

We can also demonstrate the opposite phase shifts along the $[011]$ and $[01\bar{1}]$ axes by designing unbalanced MZMs with only a single arm (SeM-MZM) meandering along the respective crystal axes, as shown in Fig.\ref{fig:DC_shift}(i, ii). The meandering arm lengths are kept identical in both devices and their nominal optical transmission spectra are similar (as shown Fig.\ref{fig:DC_shift}(a)). By parking the laser at the mid-point of the amplitude modulation spectrum (shown by the dashed lines in Fig.\ref{fig:DC_shift}(a)) and applying a DC voltage sweep of fixed polarity (\qty{0}{\volt}-\qty{32}{\volt}), we see that the differential change in modulation amplitude is opposite with DC bias. This is because the underlying MZI transmission spectrum is either red or blue detuned in the two cases, depending on waveguide orientation. Fig.\ref{fig:DC_shift}(b) plots the measured (differential) modulation amplitude, from the mid-point, as the applied DC bias is increased from 0 to \qty{32}{\volt}. The push-pull effect can clearly be seen. While the opposite nature of the effect in the two arms is easy to verify using Fig.\ref{fig:DC_shift}, the effect being exactly equal in magnitude is more challenging to quantify, given the variability between devices. We can in turn bound the difference between the two arms by quantifying the $V_{\pi}$ of the two SeM-MZM devices, which were designed to have the same meandering arm path lengths. We extract the two $V_{\pi}$ to be, respectively \qty{86}{\volt} for device (i) and \qty{93}{\volt} for device (ii).

\begin{figure}[!htbp]
    \centering
    \includegraphics[width=\linewidth]{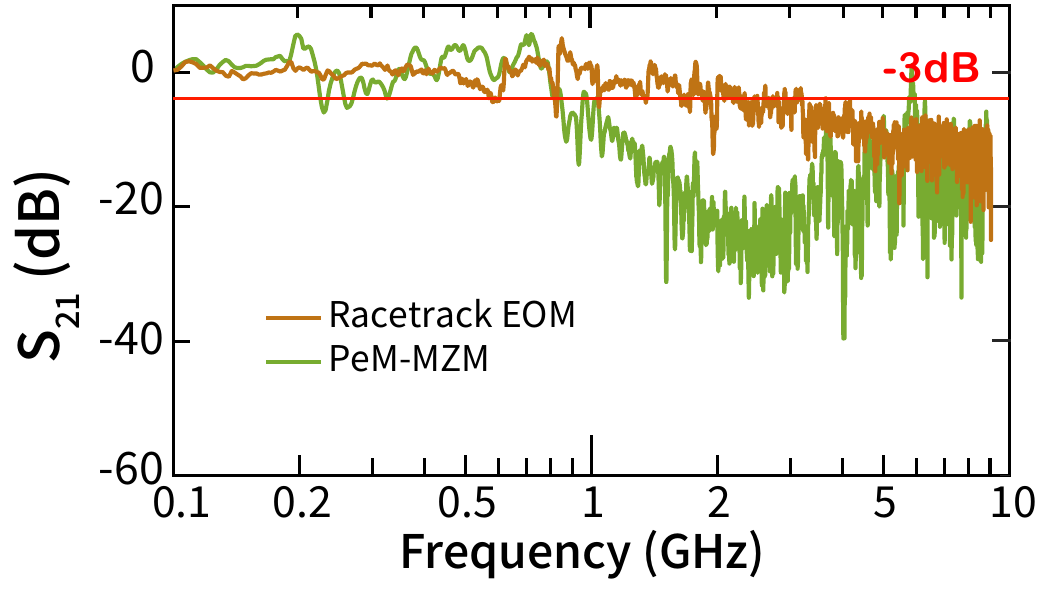}
    \caption{\label{fig:BW} Measured (normalized) EO frequency response (S\textsubscript{21}) for \qty{2}{\cm} long PeM-MZM (green) and racetrack EOM ring modulator (brown). The frequency response is normalized to \qty{100}{\mega\hertz} and the normalization procedure is discussed in Appendix \ref{sec:appendix-BW}.}
\end{figure}

We measure the modulation bandwidth (BW) of the racetrack EOM and the PeM-MZM devices using a modified version of the setup shown in Fig.\ref{fig:ss_mod}(a).  Here, we use a vector network analyzer (VNA, R$\&$S ZVL) to drive (via Port 1) the device under test with a microwave signal (\qty{0}{\dBm}, \qty{225}{\milli\volt} RMS) and sweep the modulation frequency from \qty{100}{\MHz} to \qty{9}{\GHz}. The modulated signal is measured using a high-speed amplified photodiode (Optilab, APR-10-MC) whose output is fed back into the VNA (port 2) to perform a standard EO $S_{21}$ measurement. Fig.\ref{fig:BW} plots the normalized electro-optic frequency response of the racetrack (brown) and PeM-MZM (green) devices. The device response is normalized to \qty{100}{\mega\hertz}, cf. Appendix \ref{sec:appendix-BW} for details on the normalization procedure. The extracted \qty{3}{\decibel} modulation bandwidths of the racetrack EOM and PeM-MZM devices are $\approx$ \qty{2}{\GHz} and $\approx$ \qty{0.8}{\GHz} respectively. In these proof-of-principle devices, the electrodes (see Fig.\ref{fig:schematic}(a)) were not optimized for high-speed operation, but more to ease fabrication constraints in order to demonstrate the push-pull effect in \qty{}{cm}-scale devices. Therefore, our BW is primarily limited by the RC time constant of these lumped element electrodes.

\section{Discussion}

While the results outlined in this paper clearly demonstrate the orientation dependent push-pull effect in the PeM-MZM devices, and the scale (\qty{2}{\cm} suspended arm lengths in the MZI) shows the promise of bringing MEMS based nanofabrication approaches to integrated photonics platforms, the actual device performance leaves some scope for improvement. Many of the limitations in the EOM performance metrics outlined above can be traced to conservative design choices made on the nanofabrication side to get working devices. As noted above, the scale of these devices far exceeds what has been previously demonstrated in a suspended GaAs PIC platform \cite{jiang2020suspended, khurana2022piezo}, coupled with the additional metallization constraints to generate the vertical field required at the waveguides.

Below, we outline how the various components of the PeM-MZM can be improved to achieve state-of-the-art modulator performance \cite{dagli1999wide, walker2013optimized}, keeping in mind the trade-offs between increased device complexity and reduced fabrication yield. The three main components to improve are the underlying passive optical performance (insertion and propagation loss), improving the modulation efficiency and increasing the modulation bandwidth. We consider each in turn.

While we are clearly able to demonstrate the orientation-dependent push-pull effect using the PeM-MZM devices and achieve working EOMs, the underlying passive device optical performance needs improvement. In the device shown in Fig.\ref{fig:SEM}(a), we measure an end-to-end insertion loss of \qty{29.8}{\dB}, which we can sub-divide into \qty{7.0}{\dB} per grating coupler (2$\times$), \qty{1.0}{\dB} per Y-splitter (2$\times$) and \qty{13.8}{\dB} of propagation loss. Appendix \ref{sec:appendix-opt} provides further details on the loss extraction of the individual components. The optical propagation loss of \qty{5.5}{\dB\per\cm}, extracted from the loaded quality factor of the racetrack resonators fabricated on the same chip, is  2.3$\times$ greater than the \qty{2.4}{\dB\per\cm} \cite{thomas2023quantifying} that we have demonstrated in purely passive devices before. 

The excess loss in the grating coupler is mainly due to an incomplete undercut of the underlying AlGaAs buffer layer. As noted in the fabrication procedure (Appendix \ref{sec:appendix-fab}), we rely on a timed HF acid etch to remove the AlGaAs layer and suspend the waveguides. Given the lack of tensile stress in the GaAs device layer, overetching the buffer layer causes the membranes to sag \cite{jiang2020suspended} and given the scale of the devices (\qty{2.5}{\cm} in each arm and \qty{2}{cm} suspended sections), we were keen to prevent waveguide collapse with a view towards getting functional devices. Therefore, we restricted the (over)-etch time, and that resulted in an incomplete undercut of the AlGaAs sacrificial layer with the worst affected location being the grating coupler on account of its size, more specifically, the distance from the centre of the component to the nearest etch window. With process optimization, we should be able to achieve the loss metrics we have previously demonstrated \cite{thomas2023quantifying} on these \qty{}{cm}-scale devices. Moving to wider waveguide widths ($\approx$\qty{750}{\nm}) is a potential solution as it reduces surface loss while maintaining single-mode operation, although it comes at the cost of device footprint as the minimum bend radius increases from $\approx \qty{10}{\um}$ to $\approx \qty{20}{\um}$ as the waveguide width is increased from \qty{550}{\nm} to \qty{750}{\nm}. 

\begin{figure}[H]
    \centering
    \includegraphics[width=\linewidth]{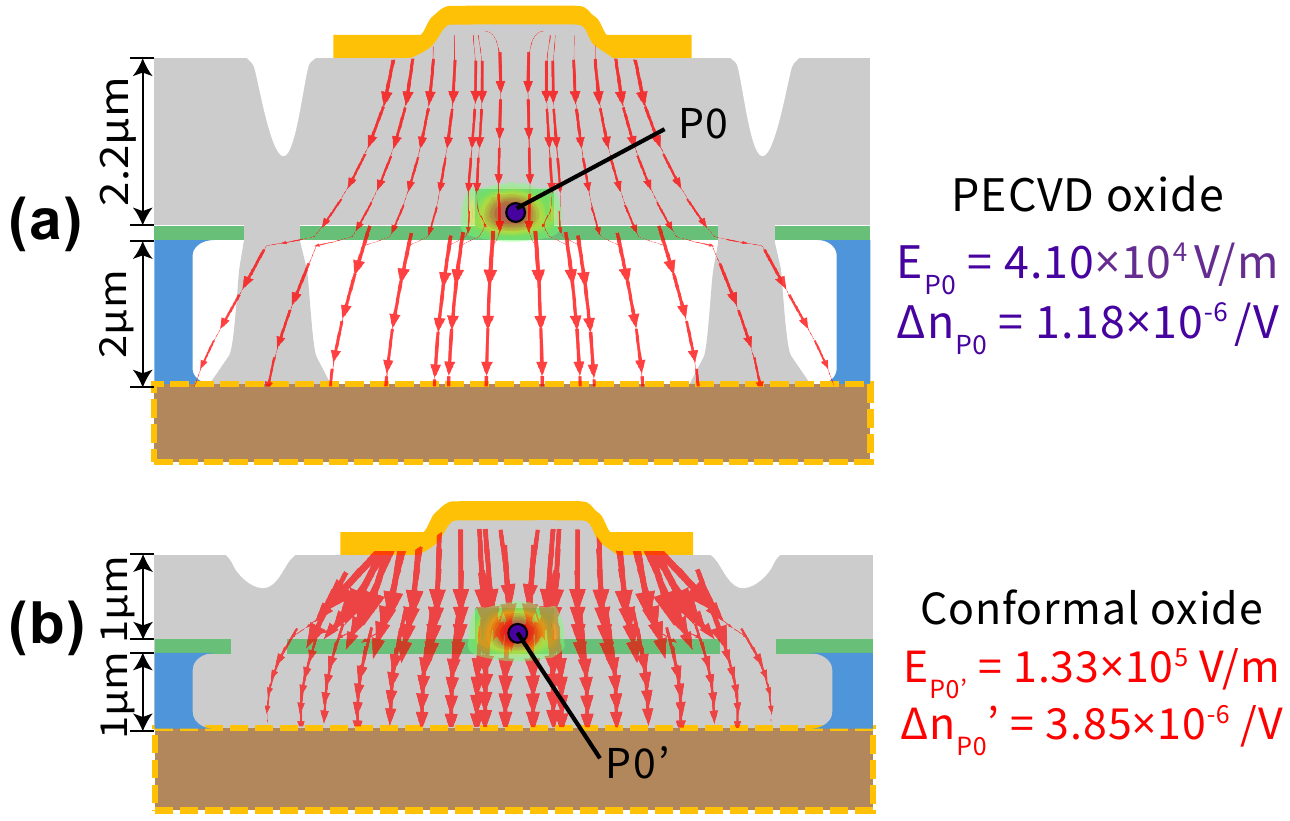}
    \caption{\label{fig:improveVpi} Electric field distribution comparison of the suspended GaAs waveguide devices shown in this work (a) with the proposed optimized geometry (b). Both the top and bottom cladding spacing to the electrodes can be reduced from $\approx$ \qty{2}{\um} in the current devices to \qty{1}{\um} without affecting optical performance. More importantly, by using conformal PECVD oxide deposition, the field strength at the waveguide (and the associated index change) can be significantly improved, as discussed in the main text. The FEM simulation of the local electric field strength is overlaid with optical mode and depicted using arrowheads that are scaled proportionally. Point P\textsubscript{0} locates the center of waveguide.}
\end{figure}

The second area of improvement, is the optimization of top and bottom cladding thickness, and electrode design to maximize the refractive index change (${\Delta}n$) per unit applied voltage and therefore maximize the modulation efficiency. In a vertical geometry like the GaAs EOM, the device can be approximated, to first order, as a series of three capacitors with dielectric constants roughly corresponding to the top cladding, waveguide and bottom cladding respectively. The voltage drop for such a series capacitor configuration scales inversely with the dielectric constant, which means a significant fraction of the field drops across the bottom air cladding. Both the top and bottom cladding thickness can be reduced by half to \qty{1}{\um} from the current devices without affecting optical performance significantly, and ensuring higher electric field strengths for a given applied voltage. By moving to a top and bottom oxide cladding using conformal PECVD \cite{abelson2020new}, we can improve the electric field strength by $\approx$ 3.3$\times$ and the overall ${\Delta}n$ by $\approx$ 3.3$\times$, cf. Fig.\ref{fig:improveVpi}. By building the same \qty{2}{\cm} PeM-MZM devices, we expect a $V_{\pi}\approx$ \qty{9.0}{\volt}. We would like to emphasize here that this optimization is performed keeping the GaAs device layer thickness fixed at \qty{340}{\nano\meter} in keeping with standard silicon photonics foundry offerings. Increasing the thickness to \qty{500}{\nano\meter} brings the $V_{\pi}$ down to $\approx$ \qty{5.5}{\volt} for similar length devices.

The final area of improvement to the devices reported in this work is incorporating travelling wave electrodes around the waveguides and velocity matching the microwave and optical fields with a view towards increasing the operational bandwidth. While the design of travelling wave electrodes is well-understood for GaAs \cite{shin2005novel, walker2013optimized}, adapting these designs to our tightly folded geometries while maintaining a low microwave insertion loss will require a re-optimization of the optical and microwave performance to maximize the device figure of merit. A second fabrication challenge that needs to be addressed is the thickness of the metal electrodes. To reduce the resistive loss at high frequencies, the metal thickness needs to be $>$ \qty{500}{\nm}, and the compatibility of such a dense metal stack with a suspended waveguide platform needs to be demonstrated in practice.  

\section{Conclusions}

We have demonstrated \textit{true} push-pull electro-optic modulators in a suspended GaAs PIC platform by exploiting the orientation induced asymmetry of the Pockels $r_{41}$ coefficient and folding the two arms of an MZI along orthogonal crystal axes ([011] and [01$\bar{1}$], respectively). We also show that sub-\qty{}{\um} mode confinement can be maintained across \qty{}{\cm}-scale devices in a suspended platform with relatively high-performance. This work provides a proof-of-principle demonstration of the idea of using geometry to exploit tensorial coefficients in crystalline media, mainly compound semiconductors, and serves as a building block for engineering quasi-phase matched interactions in curvilinear geometries in materials with $\bar{4}$ crystal symmetry \cite{kuo2014second}. By pushing on the surface loss frontier through improved surface passivation \cite{thomas2023quantifying}, these devices can potentially approach the regime of \textit{mesoscopic} nonlinear optics \cite{jankowski2024ultrafast}. As outlined in the introduction, semiconductor based EOMs have certain unique advantages over traditional ferroelectric insulators, but realizing these benefits, especially from a systems perspective, requires a coordinated effort on the photonics, microwave, materials and manufacturing fronts.

\section{Acknowledgments}

We gratefully acknowledge funding support from the UK's Engineering and Physical Sciences Research Council (GASP, EP/V052179/1) and the European Research Council (ERC-StG, SBS 3-5, 758843). Nanofabrication was carried out using equipment funded by an EPSRC capital equipment grant (QuPIC, EP/N015126/1). The GaAs wafers were sourced from the UK's national epitaxy facility in Sheffield (EP/X015300/1). We would like to thank Laurent Kling, Stephen Clements, Robert Walker, Ian Farrer, Edmund Clarke and Andrew Murray for helpful suggestions and feedback.

\appendix{}
\section{Fabrication Process and sources of excess loss}
\label{sec:appendix-fab}
\begin{figure}[H]
    \centering
    \includegraphics[width=1\linewidth]{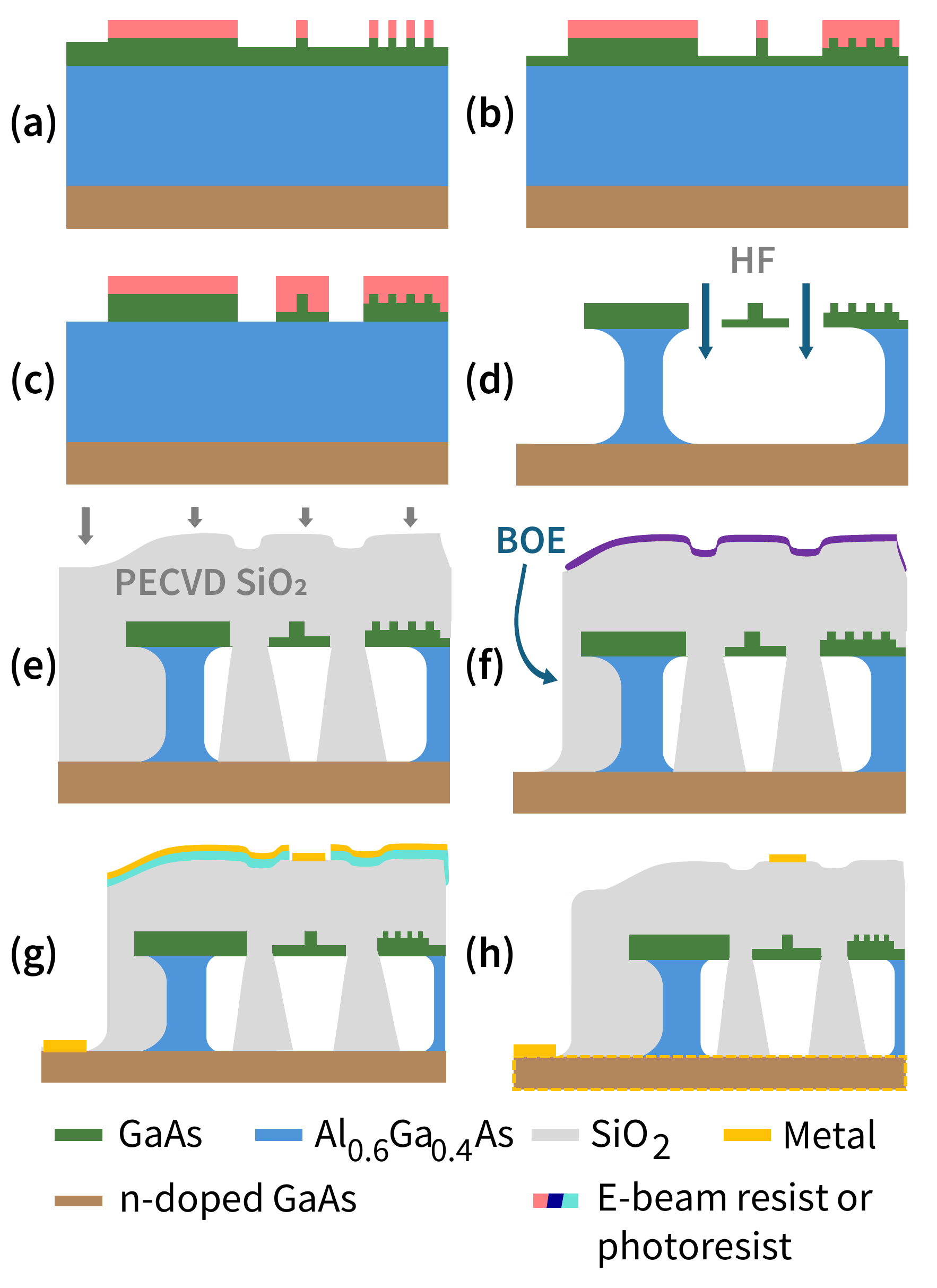}
    \caption{Schematic of the fabrication process flow used to build the suspended GaAs EOMs}
    \label{fig-fabflow}
\end{figure}

\begin{figure}[H]
    \centering
    \includegraphics[width=1\linewidth]{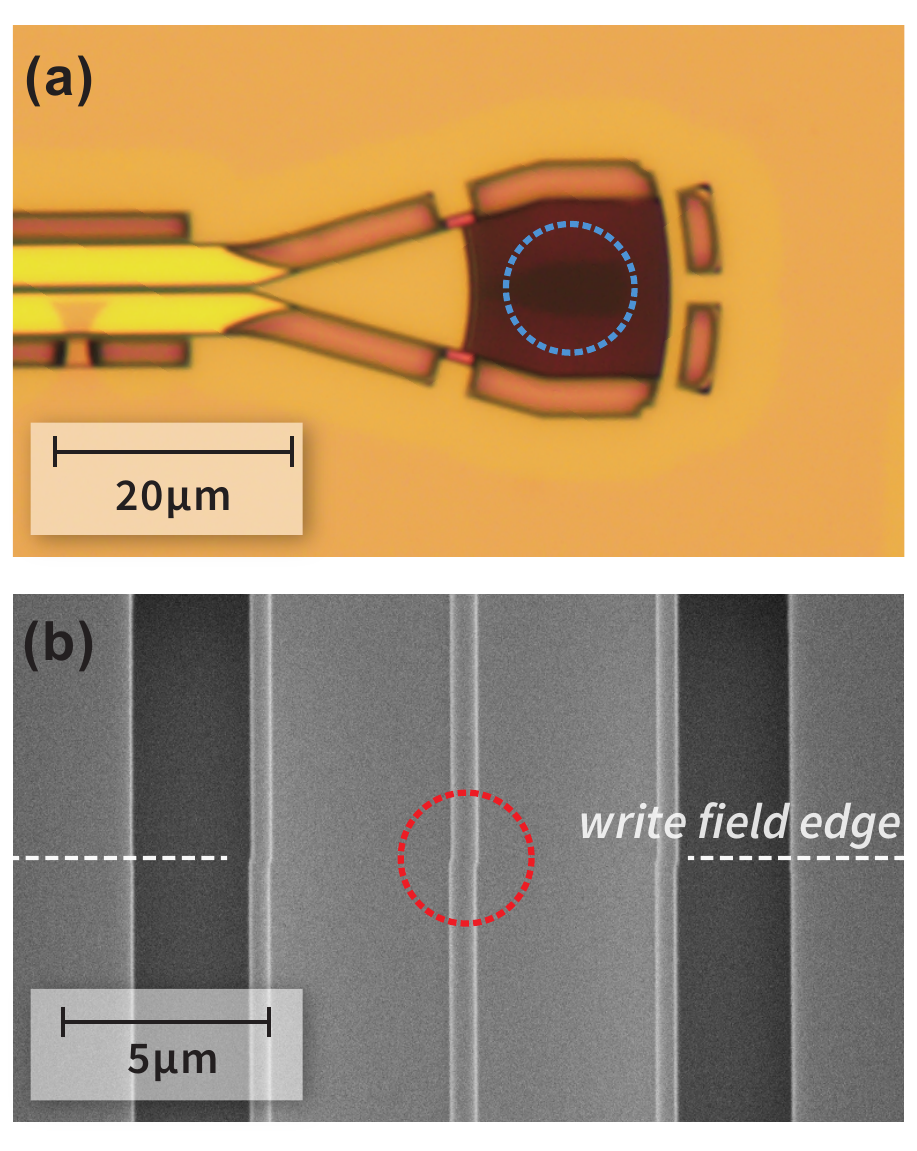}
    \caption{Examples of fabrication imperfections: (a) partially-suspended grating coupler due to an incomplete buffer layer undercut (blue dashed circle), (b) stitching error (red dashed circle) induced when EBL pattern spans cross the write field $500\times 500$\qty{}{\micro\metre} boundaries, which is inevitable in a structure the size of a PeM-MZM, see Fig.\ref{fig:layout&writefield}.}
    \label{fig:fab_imperfect}
\end{figure}

The fabrication of suspended EOMs starts with an epitaxially grown \SI{340}{nm} [100] GaAs device layer over a \qty{2}{\micro\metre} Al\textsubscript{0.6}Ga\textsubscript{0.4}As buffer layer, with an n-doped (\qty{1e18}{\per\cubic\cm}) GaAs substrate. Fig.\ref{fig-fabflow} shows the basic process flow of the fabrication, which builds on our passive device process flow \cite{jiang2020suspended}. As shown in Fig.\ref{fig-fabflow}(a-c), the photonic structures and tethers (not indicated in the cross-section)  are defined by three layers of aligned electron-beam lithography using hydrogen silsesquioxane (HSQ), and etched by inductively coupled plasma etching using an Ar/Cl\textsubscript{2} chemistry. Along with tethers, etching windows are defined around photonic devices where the GaAs device layer is fully etched through to the AlGaAs buffer. The GaAs etching rate is calibrated by measuring thickness using ellipsometer (J. A. Woollam). As shown in Fig.\ref{fig-fabflow}(d), all structures are then suspended by selectively removing the underlying AlGaAs layer through the etch windows using hydrofluoric acid (HF). The HF etch time must be carefully monitored to ensure complete suspension of the structures without overetching the AlGaAs, which could otherwise lead to membrane collapse. Fig.\ref{fig:SEM}, inset (ii), shows the waveguide after HF release, where AlGaAs is removed and GaAs substrate can be seen from top.  Fig.\ref{fig:fab_imperfect}(a) show a representative image of a grating coupler not being fully released which contributes to the excess insertion loss discussed in the main text. The GaAs surface is then passivated by depositing a \qty{7}{\nm} alumina layer via atomic layer deposition \cite{thomas2023quantifying}. 

A \qty{2.2}{\micro\metre}-thick silicon oxide (SiO\textsubscript{2}) layer is then deposited over the device using plasma-enhanced chemical vapor deposition. The SiO\textsubscript{2} enhances the structural rigidity, and its thickness must exceed that of the undercut buffer layer to enable complete filling of the etch window to provide mechanical rigidity, as shown in Fig.\ref{fig-fabflow}(e). The structural rigidity is verified by its resistance to collapse in an ultrasonic bath. In step (f), etch windows are opened in the oxide layer around the devices, and wet etching with 5\% HF is performed to expose the underlying doped GaAs substrate. A 45/15/\qty{400}{\nm} AuGe/Ni/Au metal layer is then deposited by electron-beam evaporation. To achieve an ohmic contact between metal and substrate, the devices are subsequently annealed at \qty{445}{\degreeCelsius} for \qty{1}{\minute} in a nitrogen (N\textsubscript{2}) environment. The measured resistance improves from an as-deposited \qty{1.3}{\mega\ohm} to \qty{7.34}{\ohm} after annealing, cf. Table \ref{tab:contact_resistance}.

\begin{figure}[H]
    \centering
    \includegraphics[width=1\linewidth]{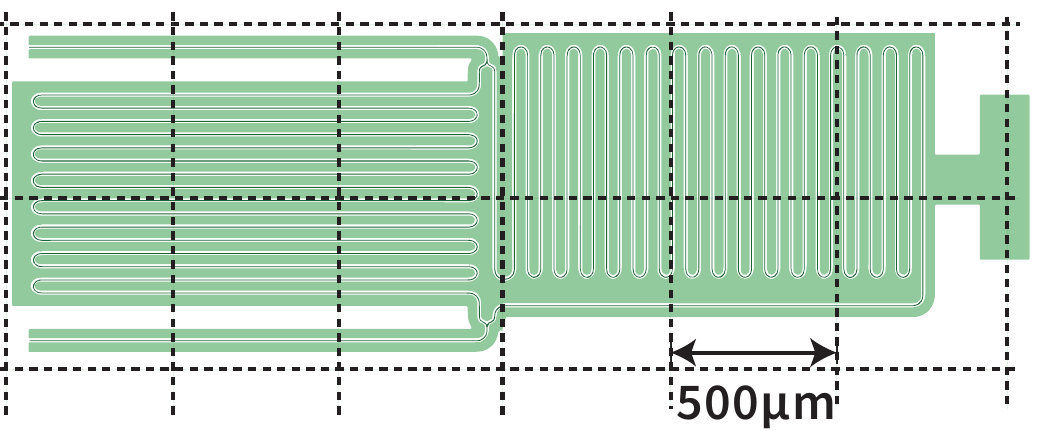}
    \caption{Layout of the 2cm PeM-MZM device (green) divided into $500\times500$\qty{}{\um} write fields (black dashed) for EBL. Waveguides that cross the write field edges are prone to stitching errors if the write field alignment procedure is not properly calibrated. These errors are magnified for a device with the scale of the PeM-MZM.}
    \label{fig:layout&writefield}
\end{figure}

In addition to the incomplete undercut issue discussed above, a second source of excess insertion loss in our experiments is down to the \qty{}{\cm}-scale of our devcies. The EBL system automatically divides the mask layout into \qty{500}{\um}$\times$\qty{500}{\um} write fields and the patterns at the write field edges are stitched together via calibrated stage movement, as shown in Fig.\ref{fig:layout&writefield}. However, the write field alignment calibration procedure is prone to various errors which lead to waveguide breaks and offsets at the write field edges. A representative example is shown in Fig.\ref{fig:fab_imperfect}(b). Given the number of write fields a pattern of the scale of the PeM-MZM must cross, even small offset errors can lead to a large cumulative insertion loss.

\begin{table}[H]
    \centering
    \renewcommand{\arraystretch}{1.25} %
    \begin{tabular}{>{\centering\arraybackslash}p{0.6\linewidth}|>{\centering\arraybackslash}p{0.4\linewidth}}
         Measured results&Resistance (\qty{}{\ohm})\\ \hline 
         Before RTA&  $(1.31\pm 0.2) \times 10^6$\\ 
         After RTA&  $7.34\pm 0.02$\\ 
    \end{tabular}
    \caption{Contact resistance comparison with/without rapid thermal annealing (RTA)}
    \label{tab:contact_resistance}
\end{table}

\section{Theoretical $\Delta n_\text{eff}$ calculation}
\label{sec:appendix-simu}

As mentioned in section II, the EO-induced refractive index change in GaAs can be expressed as:
\begin{equation}
\begin{aligned}
    \Delta n &= \Delta n_\text{LEO} +\Delta n_\text{QEO}\\&=  \pm\frac{1}{2} n^3 r_\text{41}E_{\perp(100)]} +\frac{1}{2} n^3 R_\text{21}E^2_{\perp(100)]}
    \label{eq:neff_definition}
\end{aligned}
\end{equation}

\begin{figure}[H]
    \centering
    \includegraphics[width=1\linewidth]{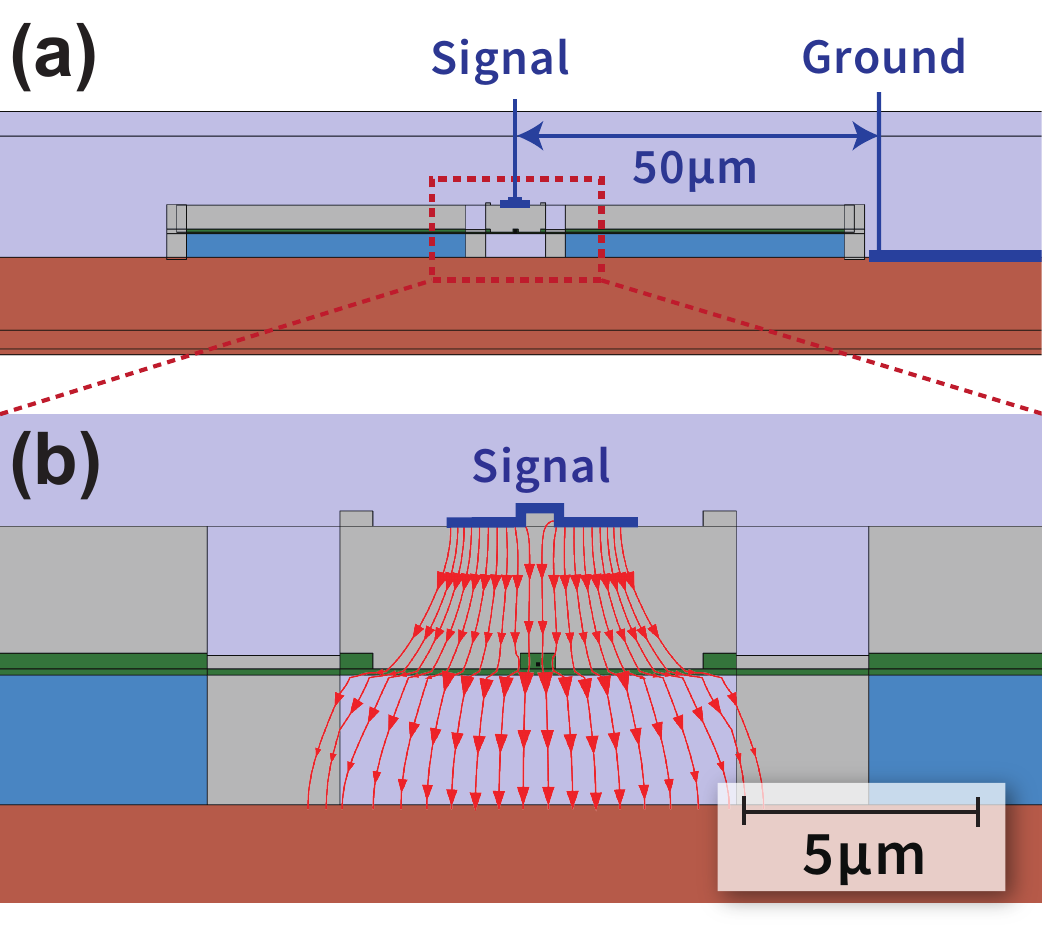}
    \caption{FEM simulation of the GaAs EOM waveguide. (a) The cross-sectional waveguide structure, showing air (light blue), SiO\textsubscript{2} (grey), GaAs (green), AlGaAs (blue) and doped-substrate (orange). (b) The red streamlines represent the distribution (local field strength) and direction of the electric field in the waveguide region. The entire doped substrate serves as a conductor and effectively grounded. The physical locations of the signal and ground contact in the actual device are indicated in (a).}
    \label{fig:simulation}
\end{figure}

for a GaAs waveguide operating at \qty{1550}{\nm} with linear EO (Pockel) coefficient  $r_\text{41} =$ \qty{-1.5e-12}{\meter\per\volt} and quadratic EO (Kerr) coefficient $R_\text{21}$ = \qty{-5.1e-21}{\square\meter\per\square\volt}. In our waveguide design, the electric field strength reaches approximately \qty{4e4}{\volt\per\meter}, and the QEO effect contribution is negligible in comparison to the linear EO effect. Thus, we approximate the EO effect using only the linear EO contribution, which is numerically calculated using an overlap integral between optical mode and the microwave electric field distribution. The calculation can be done by building the electric circuit module in COMSOL Multiphysics (v6.2), where both signal and ground terminals (Fig.\ref{fig:simulation}(a)) are treated as ideal electrical conductors. The shape of the signal terminal line in Fig.\ref{fig:simulation}(a) imitates the topography of the deposited oxide over rib waveguide. In the real device, the ground electrode is placed at the same location as shown in Fig.\ref{fig:simulation}(a), under which he GaAs substrate is set with a doping concentration of \qty{1e18}{\per\cm\cubed}. The DC electric field distribution is shown in Fig.\ref{fig:schematic}(b) and is used to calcuate the induced index change.

The effective index change of the optical TE mode in the GaAs waveguide can be calculated as \cite{dagli_high-speed_2007_ch4}:

\begin{equation}
\Delta n_\text{eff,TE} = \frac{\iint \Delta n_\text{TE} |\mathcal{E}|^2  dS}{\iint |\mathcal{E}|^2  dS} = \pm\frac{1}{2}\frac{\iint n^3 r_\text{41}E_{\perp, [100]} |\mathcal{E}|^2  dS}{\iint |\mathcal{E}|^2  dS}
\end{equation}
where $\mathcal{E}$ represents the electric field distribution of the optical TE mode and $E_{\perp}$ the vertically oriented microwave (DC) electric field component. The integration is carried out over the entire optical mode region. For a \qty{540}{\nm} wide rib waveguide with a  \qty{2}{\um}-thick top oxide and \qty{2}{\um} bottom air gap,  the calculated $\Delta n_\text{eff}$ is \qty{1.279e-6}{\per\volt}. Table \ref{tab:simulation_dneff}, also lists the experimentally extracted values for ${\Delta}n$ for the PeM-MZM and the racetrack EOM devices. We note that the optical mode in the racetrack bends will be slightly different from the straight waveguide modeshape indicated here.
\begin{table}[H]
    \centering
    \renewcommand{\arraystretch}{1.25} %
    \begin{tabular}{>{\centering\arraybackslash}p{0.5\linewidth}|>{\centering\arraybackslash}p{0.5\linewidth}}
         Result&$\Delta n_\text{eff}$ (\qty{}{\per \volt})\\ \hline 
         Simulation&  \qty{1.279e-6}{}\\ 
         Racetrack EOM&  \qty{1.086e-6}{}\\ 
         PeM-MZM&  \qty{6.972e-7}{}\\ 
    \end{tabular}
    \caption{Comparison of simulated and experimentally extracted $\Delta n_\text{eff}$ values for the racetrack EOM and PeM-MZM devices disucssed in the main text.}
    \label{tab:simulation_dneff}
\end{table}

\begin{figure}[H]
    \centering
    \includegraphics[width=\linewidth]{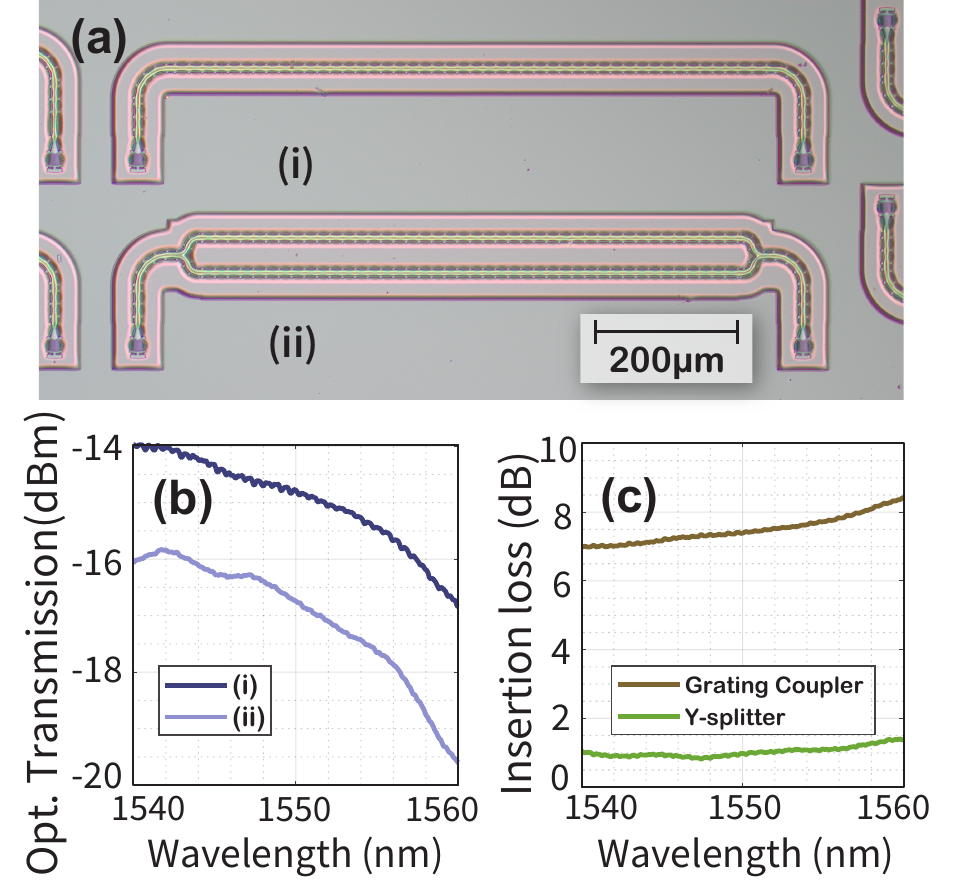}
    \caption{Optical characterization of EOM devices. (a) Microscopic images of stand-alone structures used to characterize grating couplers (i) and Y-splitters (ii). (b) Optical transmission when an optical power of \qty{0}{\dBm}(\qty{1}{\milli\watt}) is coupled into the devices. (c) Extracted insertion loss of the grating coupler (brown) and Y-splitter (green).}
    \label{fig:Micrograph-Fiber2Fiber}
\end{figure}

\section{Characterizing optical performance in GaAs EOMs}
\label{sec:appendix-opt}

The insertion losses of the passive optical components including grating couplers and Y-splitter reported in the main text are determined using the test structures shown in Fig.\ref{fig:Micrograph-Fiber2Fiber}(a). Two fiber to chip grating couplers with a pitch of \qty{889}{\micro\metre} were made, matching the fiber pitch used in the PeM-MZM device described in the main text. An optical power of \qty{0}{\dBm} (\qty{1}{\milli\watt}) from laser was coupled into the devices and the optical transmission of devices are plotted in Fig.\ref{fig:Micrograph-Fiber2Fiber}(b). Spectrum (i) reflects the total loss from two grating couplers measured on the top device in Fig.\ref{fig:Micrograph-Fiber2Fiber}(a), while spectrum (ii) includes the additional losses introduced by two Y-splitters and is measured on the bottom device in Fig.\ref{fig:Micrograph-Fiber2Fiber}(a). Using these measurements (and the assumption that the waveguide loss contribution over this length is negligible), we can extract the grating coupler insertion loss as $\approx$ \qty{7.0}{\dB} at \qty{1540}{\nm}, and the excess insertion loss of the Y-splitter is $\approx$ \qty{1.0}{\dB}. 

\begin{figure}[h] 
    \centering 
    \includegraphics[width=1\linewidth]{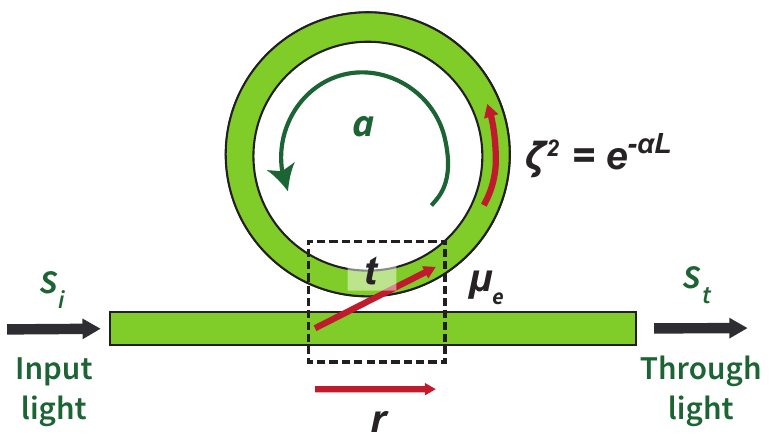}    
\caption{ \label{fig:RTmodel} Schematic of the microring resonator showing the different parameters used in the temporal coupled mode theory formalism. $s_\text{i}$ and $s_\text{t}$ denote the input and transmitted optical fields, with $|s|^2$ the power in respective mode. $a$ denotes the mode amplitude of the circulating resonator mode, normalized such that $|a(t)|^2$ is the total stored energy. The coupling parameters include the through-coupling coefficient $r$ and mutual coupling coefficient $\mu_e$.}
\end{figure}

\begin{figure}[H]
    \centering
    \includegraphics[width=\linewidth]{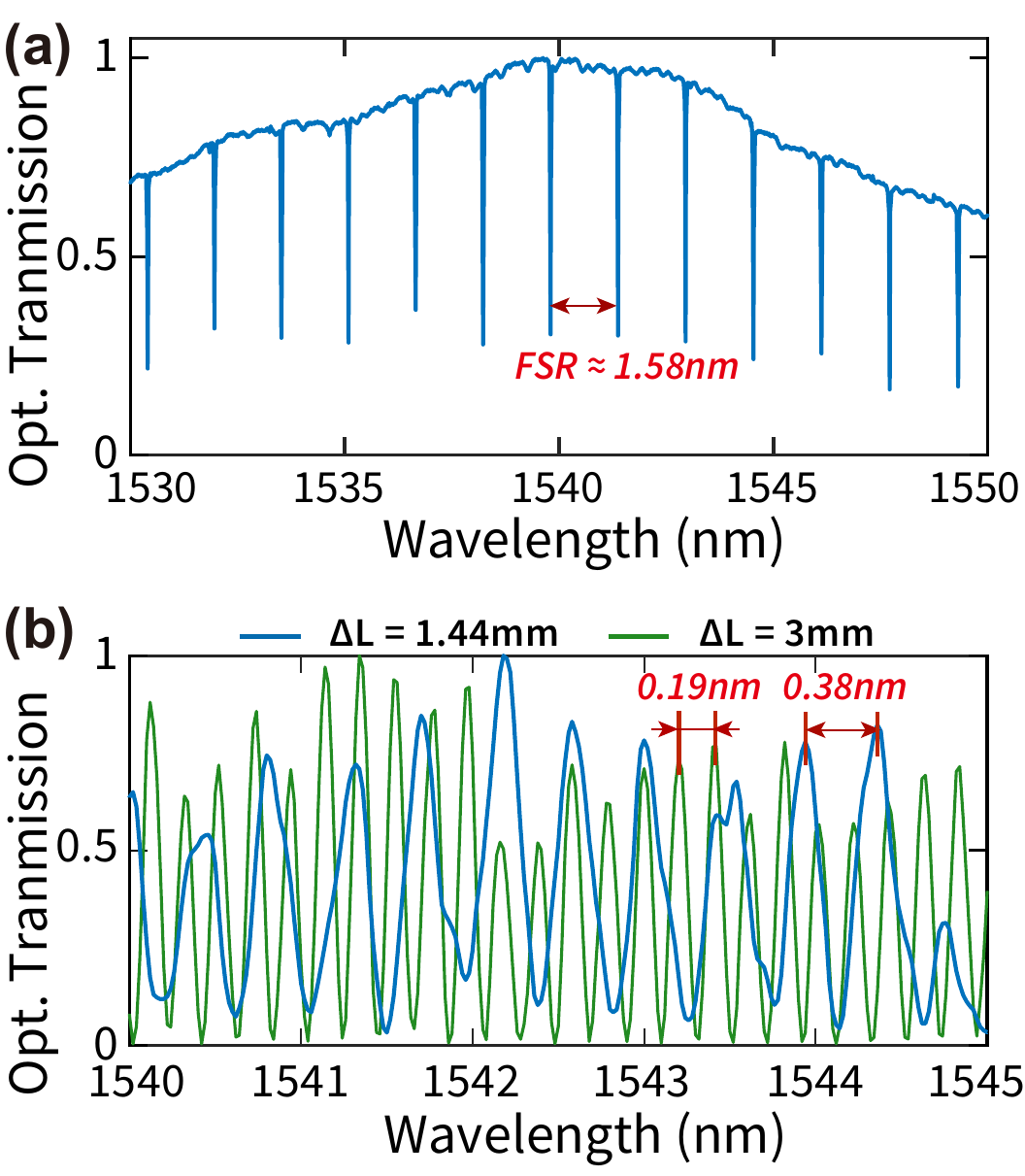}
    \caption{Normalized optical transmission spectra of GaAs EOMs measured over a wider span. (a) Racetrack resonator with optical input power of \qty{-14}{\dBm} (\qty{3.98e-2}{\milli\watt}) (b) Two PeM-MZM devices with different arm length differences (indicated in legend) with optical input power of \qty{10}{\dBm} (\qty{10}{\milli\watt})}
    \label{fig:opt_broadband_scan}
\end{figure}

To extract the intrinsic propagation loss $\alpha$ [\qty{}{\per\cm}] of GaAs waveguides, we use the intrinsic quality factor of a ring resonator as a proxy. As illustrated in Fig.\ref{fig:RTmodel} , a side-coupled resonator with a roundtrip length $L$ is characterized \cite{bogaerts_silicon_2012} by a transmission (self-coupling) coefficient $r$, a cross-coupling (waveguide to ring transfer) coefficient $t$, and a round-trip field attenuation factor $\zeta$, which is related to the propagation loss $\alpha$ by $\zeta^2 = \exp (-\alpha L)$. For simplicity, we assume the coupling is lossless, $r^2+t^2 = 1$. By measuring the resonator's loaded quality factor $Q$, extinction ratio $ER$, and free spectral range (FSR), we can determine $\alpha$, as shown below. A broadband scan of the racetrack resonator optical transmission spectrum is shown in Fig.\ref{fig:opt_broadband_scan}(a). From the measured resonances, we extract an FSR of \qty{1.58}{\nm} and fit individual resonance dips to obtain the $Q$ and $ER$. The group index ($n_g$) can be calculated as:

\begin{equation}
    n_g = \frac{\lambda_\text{res}^2}{\text{FSR}\cdot L}
\end{equation}

The measured quality factor $Q$ and extinction ratio $ER$ are related to the model parameters of Fig.\ref{fig:RTmodel} by:

\begin{equation}
\begin{aligned}
Q &= \frac{\pi n_gL\sqrt{r\zeta}}{\lambda_\text{res}(1-r\zeta)}\\
ER = \frac{T_\text{max}}{T_\text{min}}& = \frac{(r+\zeta)^2(1-r\zeta)^2}{(r-\zeta)^2(1+r\zeta)^2}
\label{eq:Q_ER}
\end{aligned}
\end{equation}

\begin{figure}[H]
    \centering
    \includegraphics[width=\linewidth]{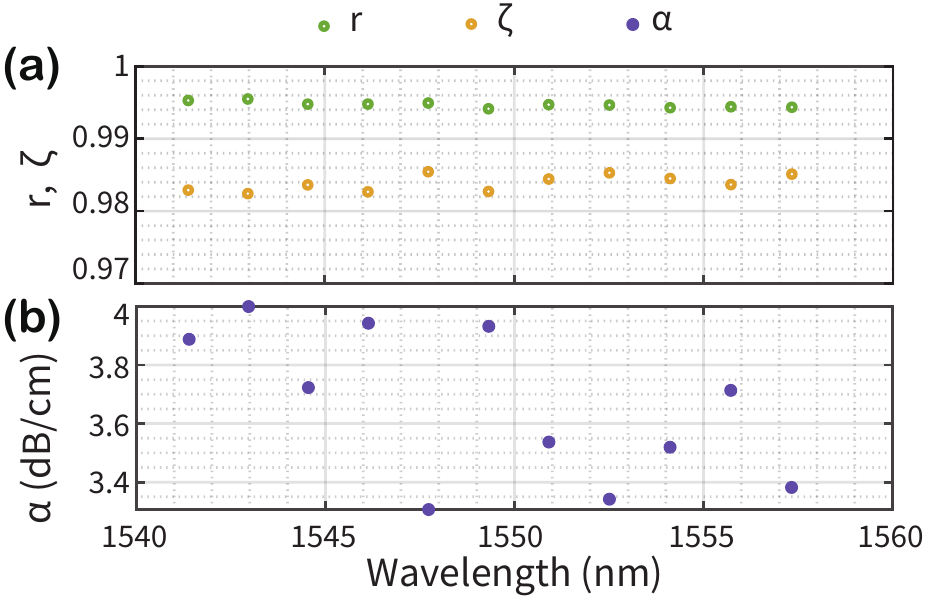}
    \caption{Extracted (a) cross-coupling coefficient $r$ and roundtrip attenuation coefficient $\zeta$, and (b) the calculated waveguide propagation loss $\alpha$ as functions of wavelength, obtained from the optical resonance of racetrack EOM in Fig.\ref{fig:opt_broadband_scan}(a).}
    \label{fig:r&zeta}
\end{figure}

Solving Eq.\ref{eq:Q_ER} using $Q = 1.467\times10^5$, extinction ratio $ER = $ \qty{4.83}{\dB} and free spectral range $\text{FSR} = $ \qty{1.58}{\nm} for the racetrack resonator yields $r \sim 0.995$ and $\zeta\sim 0.984$, confirming that the resonator operates in the under-coupled regime ($r>\zeta$). This under-coupling behavior is consistent with our prior work \cite{thomas2023quantifying}. Fig.\ref{fig:r&zeta} shows the transmission and attenuation coefficients extracted from the $Q$ and $ER$ values of the different optical resonances of the racetrack resonator. The extracted round-trip loss corresponds to a propagation loss of $\approx$ \qty{3.71}{\dB\per\cm}.

The excess losses due to random fabrication imperfections are more likely to occur in large-scale devices such as PeM-MZM which have a large and dense on-chip footprint. Given the measured transmission spectrum, it is difficult to locate and isolate (de-embed) the source of loss. Nevertheless, we can still use broadband spectral measurements to verify whether the interference length of the device matches the designed length difference of two arms. The broadband scan spectra of the PeM-MZM designed with varying arm length differences are plotted in Fig.\ref{fig:opt_broadband_scan}(b). Two types of PeM-MZMs with arm length difference ($\Delta L = L_1-L_2$) \qty{1.4}{\milli\meter} and \qty{3}{\milli\meter} are compared. The predicted fringe periods of the PeM-MZM in the two cases are \qty{0.41}{\nm} and \qty{0.20}{\nm} respectively, agreeing well with the measured spectral period of \qty{0.38}{\nm} and \qty{0.19}{nm} respectively.

\section{Evaluating Linear Electro-optic Performance of GaAs EO modulators}
\label{sec:appendix-param}

Assuming a sinusoidal modulating RF signal at frequency $\Omega_m$ is applied to an optical waveguide carrying an optical carrier at frequency $\omega_o$, the resulting phase modulated signal in the waveguide can be expressed as \cite{black_introduction_2003}:

\begin{equation}
\begin{aligned}
    E_o(t)&= E_0e^{i(\omega_ot+ \beta sin{\Omega_mt})} \\
    &= E_0\Big[A_0 J_0(\beta)e^{i\omega_ot}+A_1^+ J_1(\beta)e^{i(\omega_o+\Omega_m)t} \\
    &\quad\quad\quad\quad\quad\quad+A_1^- J_1(\beta)e^{i(\omega_o-\Omega_m)t}\Big] \\
    &\quad\quad\quad\quad\quad\quad+O\left(E_0A_n^\pm J_n(\beta)e^{i(\omega_0 \pm n\Omega_m)t}, \, n \geq 2\right)
    \end{aligned}
    \label{eq:phase_mod}
\end{equation}

Here, $\beta$ denotes the modulation index, representing the maximum phase shift induced by EO modulation. The sidebands are located at integer multiples of $\Omega_m$ relative to the optical carrier. $J_0$ , $J_n^\pm$ are Bessel functions of the first kind \cite{abramowitz1948handbook}, and $A_0$ and $A_n^\pm $ represent the spectral transfer coefficients of the optical carrier and the upper and lower sidebands which are characteristic of the device under test. For a bare waveguide, $A_i=1$. Under the small-signal approximation ($\beta\ll\pi/2$), we can ignore the contribution of the higher-order sidebands ($n\geq2$). In this regime, the lock-in signal at $\Omega_m$, which we refer to as the modulation amplitude (AM) in the main text is formed by the beating of the two sidebands with the optical carrier on the photodetector. By fitting this measured AM spectra across different device types (racetrack resonator and the MZI), the modulation index $\beta$ can be extracted, and we can quantify the tunability of the modulator. In the following sections, we detail the procedure for extracting the modulation index from the measured optical and AM spectra of the racetrack EOM and the PeM-MZM devices respectively.

\subsection{Racetrack EOM electro-optic transfer model}

The schematic of a typical ring resonator is shown in Fig.\ref{fig:RTmodel}, which possesses a series of optical resonances at angular frequencies expressed as:
\begin{equation}
\omega_\text{r} = m\frac{2\pi c}{n_\text{eff}L}
\end{equation}
where $m$ is the mode number, $c$ is the speed of light, $n$\textsubscript{eff} is the effective index of the optical waveguide and $L$ is the round-trip cavity length of the resonator. Assuming a sinusoidal modulating RF signal at angular frequency $\Omega_m$ is applied to the optical waveguide, the refractive index varies as:
\begin{equation}
    n = n_o+\Delta n \cos{\Omega_mt}
\end{equation}
where $n_o$ is the bare GaAs refractive index and $\Delta n$ denotes the index perturbation. This index modulation can be mapped to an equivalent modulation to optical frequency:
\begin{equation}
\begin{aligned}
    \omega(t) &= \omega_{o}+\Delta\omega\cos{\Omega_mt} \\
    & = \omega_o + \Delta\omega \frac{e^{j\Omega_mt}}{2}+ \Delta\omega \frac{e^{-j\Omega_mt}}{2}
\end{aligned}
\end{equation}
with $\Delta\omega=\omega{\Delta}n/n_0$ being the peak frequency shift. The modulation index $\beta$ quantifies the induced phase shift and can be calculated as:

\begin{equation}
 \label{eq:phase_definition}
\begin{aligned}
  \phi(t) &= \int_0^t \omega(\tau)\, d\tau = \int_0^t \left[\omega_0 + \Delta\omega \cos(\Omega_m \tau)\right] d\tau \\
       &= \omega_0 t + \beta \sin(\Omega_m t), \quad \beta = \frac{\Delta\omega}{\Omega_m}
   \end{aligned}
\end{equation}

 Let $a(t)$ [\si{\joule^{1/2}}] represent the amplitude of the circulating cavity mode inside in the ring resonator depicted in Fig.\ref{fig:RTmodel}. The mode amplitude is normalized such that $|a(t)|^2$ corresponds to the total optical energy \qty{}{\joule} stored within the cavity. The frequency-modulated intra-cavity field, under the small signal approximation, can be written as:

\begin{equation}
    a(t) = a_{c}e^{j\omega_0t}+a_{+} e^{j(\omega_o +\Omega_m)t}+a_{-}e^{j(\omega_o-\Omega_m)t}
    \label{eq:a(t)}
\end{equation}
where $\omega_o$ denotes the optical carrier frequency and $\Omega_m$ is angular modulation frequency. The coefficients $a_c$, $a_+$ and $a_-$ denote the complex transfer functions of optical carrier, upper sideband ($\omega_0+\Omega_m$), and lower sideband ($\omega_0-\Omega_m$) respectively. They are determined by the spectral response of the cavity. The relationship between the intra-cavity field and the field in the coupling waveguide is determined using temporal coupled mode theory \cite{pile2014smallsignalresonator}.  To derive the coupled mode equations, we model the microring resonator as a lumped oscillator at frequency $\omega_\text{r}$ [rad/s], yielding:

\begin{equation}
    \frac{da(t)}{dt} = \left(j\omega_r-\frac{1}{\tau}\right)a(t)-j\mu_es_\text{i}(t)
    \label{eq:CMT1}
\end{equation}

\begin{equation}
    s_\text{o}(t) = s_\text{i}(t) -j\mu_ea(t) 
     \label{eq:CMT2}
\end{equation}

where $\tau$ [s] is the field decay time constant of the resonator, which characterizes the exponential decay of the intracavity field as $a(t)\propto e^{-t/\tau}$. The corresponding photon lifetime is given by $\tau_p =\tau/2 = Q/\omega_\text{res}$ and can be calculated via a measurement of the loaded quality factor $Q$ of the resonator (see Appendix \ref{sec:appendix-opt} below).  The mutual coupling coefficient $\mu_e$ [\unit{s^{-1/2}}] quantifies the coupling between the input wave $s_\text{i}(t)$ and the intracavity field $a(t)$. Its numerical value is related to the transmission coefficient $r$ (cf. Fig.\ref{fig:RTmodel}) by:

\begin{equation}
    \mu_e = \sqrt{ -\frac{2\,c\log_e|r|}{ n_g L}}
\end{equation}

where $n_g$ is the group index extracted from the measured cavity free spectral range as shown in Appendix \ref{sec:appendix-opt}. The input and output optical fields $s_\text{i}(t)$ and $s_\text{o}(t)$ in the waveguide are normalized such that $|s_i(t)|^2 = P_\text{in}$ and $|s_o(t)|^2 = P_\text{out}$ represent the input and output optical power in the waveguide in \qty{}{\watt}. Using Eq.\ref{eq:a(t)} as an ansatz for solving the coupled mode equations (Eq.\ref{eq:CMT1}, \ref{eq:CMT2}) and equating the coefficients by frequency \cite{pile2014smallsignalresonator} we get:

\begin{equation}
    \begin{aligned}
        a_c & =\frac{-j\mu_e\sqrt{P_\text{in}}}{j\delta+1/\tau}\\
        a_+ & = \frac{j}{2}\frac{a_c}{j(\delta+\Omega_m)+1/\tau}\Delta\omega_r\\
        a_- & = \frac{j}{2}\frac{a_c}{j(\delta-\Omega_m)+1/\tau}\Delta\omega_r
    \end{aligned}
\end{equation}
where $\delta = \omega-\omega_{r}$ is the detuning of the laser frequency relative to the cavity resonance and $\Delta\omega_r$ represents the maximum frequency shift of the cavity resonance under perturbation. To simplify notation, we define a generalized Lorentzian cavity response function:

\begin{equation}
    \mathcal{L}(\Omega ) = \frac{1}{j(\delta+\Omega)+1/\tau}
\end{equation}

We can then construct the output optical field amplitude as:

\begin{equation}
    \begin{aligned}
        s_\text{o}(t) & = s_\text{i}(t)-j\mu_ea(t)\\
        & = s_\text{i}(t)-j\mu_e\big(a_ce^{j\omega_ot }+ a_+e^{j(\omega_o+\Omega_m)t}+ a_-e^{j(\omega_o+\Omega_m)t}\big)\\
        & = \sqrt{P_\text{in}}e^{j\omega_ot}-j\mu_e\Big[\frac{-j\mu_e\sqrt{P_\text{in}}}{j\delta+1/\tau}e^{j\omega_ot }\\
        &\quad\quad\quad+\frac{ja_c\Delta\omega_r/2}{j(\delta+\Omega_m)+1/\tau}e^{j(\omega_o+\Omega_m)t}\\
        &\quad\quad\quad+\frac{ja_c\Delta\omega_r/2}{j(\delta-\Omega_m)+1/\tau}e^{j(\omega_o-\Omega_m)t}\Big]\\
        &= \sqrt{P_\text{in}} e^{j\omega_\text{o}t} \Big[(1-\mu_e^2\mathcal{L}(0) \\&\quad\quad\quad-\frac{j}{2}\mu_e^2\Delta\omega_r\mathcal{L}(0)\mathcal{L}(\Omega_\text{m}) e^{j\Omega_\text{m}t}\\
        &\quad\quad\quad-\frac{j}{2}\mu_e^2\Delta\omega_r\mathcal{L}(0)\mathcal{L}(-\Omega_\text{m}) e^{-j\Omega_\text{m}t} \Big]
    \end{aligned}
\end{equation}

The output optical intensity, detected by the photodetector, is therefore:
\begin{equation}
    P_\text{out} = |s_\text{o}|^2 
\end{equation}

$P_\text{out}$ thus consists of a DC term, carrier–sideband beat notes at $\pm\Omega_m$, and higher order terms $\pm2\Omega_m$. The lock-in measurement extracts the signal at $\pm\Omega_m$, which is given by:

\begin{equation}
\begin{aligned}
    P_\text{out}(\pm\Omega_\text{m})  =2 \Re\Big\{jP_\text{in}&\frac{\Delta\omega_r}{2}\mu_e^2\Big[\mathcal{L}(0)^*\mathcal{L}(-\Omega_\text{m})^* -\mathcal{L}(0)\mathcal{L}(\Omega_\text{m})\\
    &+\mu_e^2|\mathcal{L}(0)^2|(\mathcal{L}(\Omega_\text{m})-\mathcal{L}(-\Omega_\text{m})^*\Big] \Big\}
\end{aligned}
\end{equation}

The detected microwave voltage is therefore:
\begin{equation}
V_\text{out} = R_\text{PD} P_\text{out}(\pm\Omega_\text{m})
\end{equation}

where $R$\textsubscript{PD} [\unit{\volt\per\watt}] is the responsivity of the amplified photodiode. From the measured $V_\text{out}$, we can extract the corresponding frequency shift ($\Delta\omega_r$) and the equivalent wavelength shift $\Delta\lambda$ [\unit{\nm}] induced by a modulating voltage signal of amplitude $V$\textsubscript{p}. The electro-optic tunability $\eta$ is then defined as:

\begin{equation}
\eta = \frac{\Delta\lambda}{V_p} \quad[\si{\pico\meter\per\volt}]
\end{equation}

The half-wave voltage $V_\pi$ of the microring EOM is expressed as:
\begin{equation}
    V_\pi = \frac{\Delta\lambda_\pi}{\Delta\lambda}\cdot V_\text{p} \quad[\si{\volt}]
\end{equation}
with $\Delta\lambda_\pi$ being the wavelength difference between the resonator transmission minima and maxima. The induced effective index change due to the applied field is:
\begin{equation}
    \Delta n_\text{eff} = \frac{\eta}{L} \quad [\si{\per\volt}]
\end{equation}

\subsection{MZM electro-optic transfer model}
\begin{figure}[h] 
    \centering 
    \includegraphics[width=\linewidth]{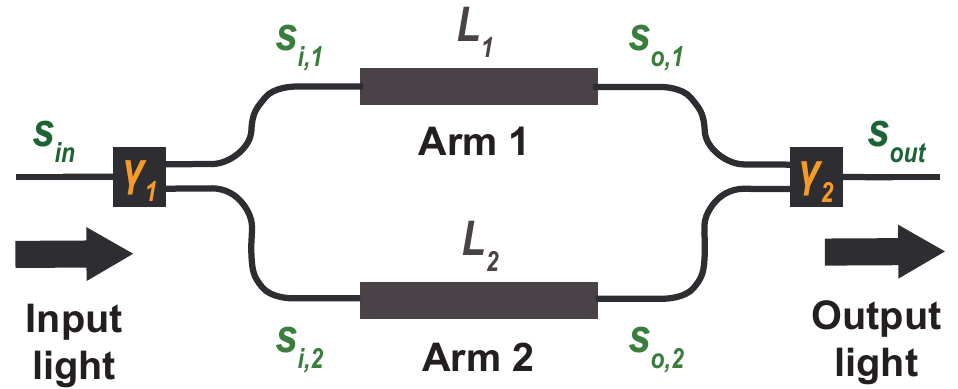}    
\caption{ \label{fig:MZMmodel} The diagram of a typical Mach-Zehnder modulator. }
\end{figure}

As shown in Fig.\ref{fig:MZMmodel}, a Mach-Zehnder interferometer (MZI) modulator splits the input light $s_\text{i}$ (defined as the nomrlized input amplitude with $|s_i|^2 = P_\text{in}$ into two branches, where phase modulation is applied separately along two arms before the optical signals recombine to generate constructive or destructive interference. Each arm introduces phase modulation that results in the generation of optical sidebands, as indicated by Eq.\ref{eq:phase_mod}. Unlike the ring-resonator based modulators - where the sidebands amplitudes are governed by cavity response (Eq.\ref{eq:a(t)}) - the MZI produces sidebands with equal amplitude, i.e. $|A_1^+| = |A_1^-|$ in Eq.\ref{eq:phase_mod}. 

We define the power splitting ratios at the input and output Y-junctions (cf. Fig.\ref{fig:MZMmodel}) as  $\gamma_{i,q}$ and $\gamma_{o,q}$, where $q = 1,2$ denotes the two arms of the interferometer. These ratios satisfy (the insertion loss can be subsumed into the overall propagation loss):
\begin{equation}
\gamma_{i,1} + \gamma_{i,2} = 1 , \quad \gamma_{o,1}+\gamma_{o,2}=1
\end{equation}

Assuming the input optical field is: 
\begin{equation}
    s_\text{in}(t) =\sqrt{P_\text{in}}e^{j\omega_ot}
\end{equation}
where $P_\text{in}$ is the input optical power in units of Watt. The field entering each arm becomes:
\begin{equation}
    s_{i,q}(t)= \sqrt{\gamma_{i,q}P_\text{in}}e^{j\omega_ot}, \quad\quad q = 1,2
    \label{eq:MZM_input}
\end{equation}

After propagation through arm $q$ with physical length $L_q$, the output field at each arm is given by:
\begin{equation}
    s_{o,q}(t)= \sqrt{\gamma_{i,q}P_\text{in}}e^{j\omega_ot}e^{(-jkL_q-\alpha L_q/2)}
\end{equation}
where $k= 2\pi n_\text{eff}/\lambda$  [\si{\radian\per\meter}]is the phase constant of the unmodulated waveguide, $\lambda$ is the optical wavelength, $n_\text{eff}$ is the effective refractive index, and $\alpha$ [\si{\per\meter}] is the optical propagation loss coefficient. The total output field after recombination is:
\begin{equation}
    s_\text{out}(t) = \sqrt{\gamma_{o,1}}s_{o,1}(t)+\sqrt{\gamma_{o,2}}s_{o,2}(t)
\end{equation}

The optical transmission of the MZI is described as:
\begin{equation}
\begin{aligned}
    P_\text{out}  & = |s_\text{out}|^2 \\
    & = P_\text{in}\Big[\gamma_{o,1}\gamma_{i,1}e^{-\alpha L_1}+\gamma_{o,2}\gamma_{i,2}e^{-\alpha L_2} \\
    &  +2\sqrt{\gamma_{o,1}\gamma_{i,1}\gamma_{o,2}\gamma_{i,2}} e^{-\frac{\alpha(L_1+L_2)}{2} } \cos\Big({\frac{2\pi n_\text{eff}}{\lambda}\Delta L}\Big)\Big]
    \end{aligned}
\end{equation}
where $\Delta L = L_1-L_2$ is the arm length difference. To evaluate how phase modulation affects the output, we consider the gradient of the output optical power:

\begin{equation}
\begin{aligned}
\frac{dP_\text{out}}{d\lambda} 
=&-\sqrt{\gamma_{o,1}\gamma_{i,1}\gamma_{o,2}\gamma_{i,2}} P_{\text{in}}  e^{-\frac{\alpha}{2} (L_1+L_2)} \\
&\quad\frac{ 4\pi n_{\text{eff}}\Delta L}{\lambda^{2}}
\sin\left(\frac{2\pi n_\text{eff}}{\lambda} \Delta L\right)
\end{aligned}
\label{eq:dPout_dwvl}
\end{equation}

If we make the assumption that the splitting ratios and any insertion loss the signal encounters in propagating the MZI are wavelength independent, then we can use the measured gradient of the transmitted optical spectrum of the PeM-MZM to determine the overall modulation efficiency. 

When a sinusoidal RF signal is applied at frequency $\Omega_m$ [\si{\radian\per\meter}], the refractive index in both arms is modulated via the electro-optic effect. The index modulation can be expressed as $ n_q(t) = n_\text{o} +\Delta n_q \sin{\Omega_mt}$, where $n_\text{o}$ is the unmodulated material index and $\Delta n_q$ is the modulation-induced index change. In the absence of modulation, the optical phase delay accumulated in each arm is simply $\phi_q^{(0)} = kL_q$. When modulation is applied, an additional time-varying phase term is introduced: 
\begin{equation}
\begin{aligned}
\phi_q &= e^{-jk_q L} \\
& = e^{-j(k_oL_q+\frac{2\pi}{\lambda}\Delta n_q L_{m,q}\sin{\Omega_mt})}\\
& = e^{-jk_oL_q}e^{-j\beta_q\sin{\Omega_mt}}
\label{eq:MZM_phi}
\end{aligned}
\end{equation}
where $L_{m,q}$ is the effective modulation length in arm $q$, and the modulation index is defined as $\beta_q = 2\pi\Delta n_qL_{m,q}/\lambda$ [\si{\radian}]. The output field from each arm can then be expanded into a series of Bessel functions \cite{abramowitz1948handbook}. Following the general Bessel expansion introduced in Eq.\ref{fig:EOmod}, the modulated output field in each arm can be expressed using the small-signal approximation ($\beta_q \ll \pi/2$) by retaining only the first-order sidebands:
\begin{equation}
\begin{aligned}
        s_{o,q}(t) & = s_{i,q}(t)e^{-\alpha L_q}\phi_q \\
        & = s_{i,q}(t)e^{-\alpha L_q}e^{-jk_oL_q}e^{-j\beta_q\sin{\left(\Omega_mt\right)}}\\
        & =s_{i,q}(t)e^{-\alpha L_q}e^{-jk_oL_q} \Big[ J_0( \beta_q) e^{j (\omega_o t)} \\
   & \quad  - J_1( \beta_q) e^{j (\omega_o + \Omega_m)t} + J_1( \beta_q) e^{j (\omega_o - \Omega_m)t}\Big] 
\end{aligned}
\label{eq:Eo_phase_mod}
\end{equation}

where $ s_{i,q}(t)$ is the input field amplitude in arm $q$, previously defined in Eq.\ref{eq:MZM_input}. The output optical field arriving at the photodiode is given by:

\begin{equation}
\begin{aligned}
    P_\text{out}(t)& = |s_\text{out}(t)|^2\\
    &= (\sqrt{\gamma_{o,1}}s_{o,1}(t)+\sqrt{\gamma_{o,1}}s_{o,2}(t))^2 \\
    & = \gamma_{o,1} |s_{o,1}(t)|^2 + \gamma_{o,2}|s_{o,2}(t)|^2\\
    & \quad + \sqrt{\gamma_{o,1}\gamma_{o,2}} [s_{o,2}^*(t) s_{o,1}(t) + s_{o,1}^*(t) s_{o,2}(t)]\\
    & = \gamma_{o,1}\gamma_{i,1} e^{-\alpha L_1-j2 k_o L_1} P_{in}[J_{0,1}^2 +   \\
    & \quad\quad 2J_{1,1}^2- 2J_{1,1} \cos{2\Omega_m t}]\\
    &\quad +\gamma_{o,2}\gamma_{i,2} e^{-\alpha L_2-j2 k_o L_2} P_{in}[J_{0,2}^2   \\
    & \quad\quad+ 2J_{1,2}^2- 2J_{1,2} \cos{2\Omega_m t}] \\
    & \quad +\sqrt{\gamma_{o,2}\gamma_{i,2}} \sqrt{\gamma_{o,1}\gamma_{i,1}} |s_o|^2 e^{-\frac{\alpha}{2} (L_1+L_2)} \\
    &\quad\quad\,\{ \cos{ k_o\Delta L}[J_{0,1} J_{0,1}+ 2 J_{1,2} J_{1,1} \\
    & \quad\quad- 2 J_{1,2} J_{1,1} \cos{2\Omega_m t} ]  + 2 \sin{ k_o\Delta L}\\
    & \quad\quad[( J_{0,1} J_{1,1} - J_{1,2} J_{0,1} ) \sin {\Omega_m t} ]\}
\end{aligned}
\label{eq:MZM_Pout}
\end{equation}

For notational clarity, we define: 
\begin{equation}
J_{0,q} = J_0(\beta_q), \quad J_{1,q} = J_1(\beta_q), \quad q = 1,2
\end{equation}

At the photodetector, the modulation amplitude signal at $\Omega_m$ arises from the beat note between the carrier and first-order sidebands. Substituting the expanded fields, the detected AM voltage is:

\begin{equation}
\begin{aligned}
    V_\text{out} &=R_\text{PD}P_\text{out}\\
    &= 4  R_\text{PD} \sqrt{\gamma_{o,2}\gamma_{i,2}\gamma_{o,1}\gamma_{i,1}} P_\text{in} e^{-\frac{\alpha}{2}(L_1+L_2)} \\
    &\quad\quad\quad\quad\quad\mathcal{J} \sin\Big(\frac{2\pi n_\text{eff}}{\lambda} \Delta L\Big)
\end{aligned}
    \label{eq:Vout_init}
\end{equation}
where $R_\text{PD}$ [\si{\volt\per\watt}]is the phodiode responsivity and $\mathcal{J}$ quantifies the net modulation amplitude from both arms:
\begin{equation}
\begin{aligned}
    \mathcal{J} &= J_{0,2}J_{1,1}-J_{0,1}J_{1,2} \\
    &= J_0(\beta_2)J_1(\beta_1)- J_0( \beta_1)J_1(\beta_2)
    \end{aligned}
\end{equation}

which captures the modulation-induced imbalance between the two arms. When  arms with equal modulation length experience identical modulation, i.e., $\beta_1=\beta_2$, the modulated phase term cancel out, resulting in $\mathcal{J}$ and thus zero AM output. Therefore, in this paper, the presence of a measurable AM signal in PeM-MZM itself demonstrates that the optical phase modulation in the two arms is anti-symmetric ($\beta_1 = -\beta_2$), i.e., acting under the push-pull configuration.

By comparing Eq.\ref{eq:dPout_dwvl} and Eq.\ref{eq:Vout_init}, the final relationship between the measured AM voltage and the gradient of the optical transmission spectrum can be expressed as: 
\begin{equation}
\begin{aligned}
     V_\text{out}&  = \Big|\frac{\lambda^2 R_\text{PD}\mathcal{J}}{\pi n_\text{eff}\Delta L} \frac{dP_\text{out}}{d\lambda}\Big|
     \end{aligned}
     \label{eq:MZM_final}
\end{equation}

From the measured $V_{\text{out}}$ and known spectral slope $dP_\text{out}/d\lambda$ [\si{\watt\per\meter}], we can extract the modulation index $\mathcal{J}$ and subsequently determine the modulation index $\beta$ and the change of effective index $\Delta n_\text{eff}$. The half-wave voltage ($V_\pi$) of the push-pull MZM model can be expressed as:
\begin{equation}
V_\pi = \frac{\lambda}{2 \Delta n_\text{eff} L_m}.
\end{equation}
where $L_m$ is the effective modulation length. Based on this, the electro-optic (EO) tunability is extracted as:
\begin{equation}
\eta = \frac{\Delta\lambda}{V_p} = \frac{1}{V_p} (\frac{\lambda}{n_\text{eff}}-\frac{\lambda}{ n_\text{eff} + \Delta n_\text{eff}})
\end{equation}

\section{Bandwidth data normalization}\
\label{sec:appendix-BW}

In Fig.\ref{fig:BW}, the normalized bandwidth for PeM-MZM and racetrack EOM are measured at the quadrature operation point, corresponding to the peaks of AM spectra (see Fig.\ref{fig:EOmod}(b,c)). During the S\textsubscript{21} measurement, a high-gain avalanche photodiode (APD) is used to amplify the  signal to improve the overall signal-to-noise ratio for the vector network analyzer (VNA) measurement. However, the measured bandwidth on the VNA now reflects the combined frequency response of both DUT and the APD. 

\begin{figure}[H]
    \centering
    \includegraphics[width=1\linewidth]{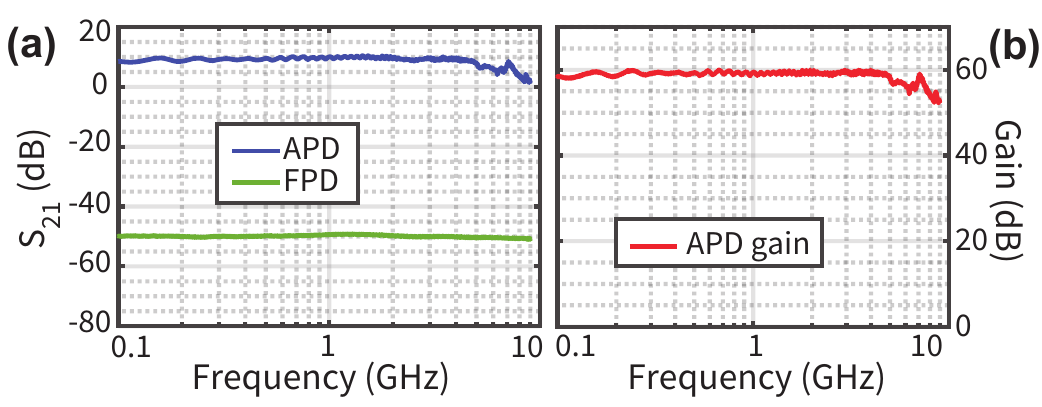}
    \caption{(a) Measured frequency response of a broadband intensity modulator using a fast photodiode (FPD, green) and the APD (blue), measured over a frequency range of \qty{100}{\GHz} to \qty{9}{\GHz} with an optical input power of \qty{-6}{\dBm} at the photodiode. (b) Extracted APD gain by comparing responses with and without APD, which is subsequently used to normalize the measured S\textsubscript{21} response of the DUT.}
    \label{fig:rawBW}
\end{figure}

To de-embed the intrinsic response of the DUT, we first characterize the gain response of the APD over the same frequency range.  This is done by using a fast photodiode (DC–\qty{30}{\GHz}, Thorlabs DXM30AF) to measure the frequency response of a broadband intensity modulator (\qty{40}{\GHz}, Thorlabs LNA6112). The input optical power is \qty{6}{\dBm} (\qty{3.98}{\milli\watt}) and the microwave frequency is swept from \qty{100}{\MHz} to \qty{9}{\GHz} with the power set to \qty{0}{\dBm}. The power incident on the photodiode was maintained constant at (\qty{-6}{\dBm}, \qty{0.25}{\milli\watt}) across the FPD and APD measurements to ensure that the extracted APD gain reflects only the detector’s frequency-dependent response and remove any dependence of the measured gain on the optical intensity. The combined S\textsubscript{21} [\qty{}{\dB}] response of the intensity modulator with and without APD is shown in Fig.\ref{fig:rawBW}(a), and the gain contribution of the APD is extracted by subtraction (in \qty{}{\dB}), as shown in Fig.\ref{fig:rawBW}(b). Using the measured APD gain spectrum, we can extract the intrinsic frequency response of the DUT as: 
\begin{equation}
    S_{21,\text{DUT}}(f)[\si{\dB}]= S_{21,  \text{DUT}+ \text{APD}}(f)[\si{\dB}]- S_{21,  \text{APD}}(f)[\si{\dB}]
\end{equation}

The normalized frequency response shown in the main text is plotted by subtracting the S\textsubscript{21} value at \qty{0.1}{\GHz} from the entire curve, in effect setting the \qty{100}{\mega\hertz} response as the unity gain reference for measuring the \qty{3}{\dB} bandwidth.

\begin{equation}
\begin{aligned}
    S_{21,\text{DUT},\text{normalized}}(f)[\si{\dB}] = &S_{21,\text{DUT}}(f)[\si{\dB}] - \\&S_{21(,\text{DUT})}(f_{\qty{0.1}{\GHz}})[\si{\dB}]
\end{aligned}
\end{equation}

\bibliography{References}

\end{document}